%% file: arxiv-2.tex
\documentclass[12pt]{article}
{\begin{Sbox}\begin{minipage}}%
{\end{minipage}\end{Sbox}\fbox{\TheSbox}}%

\usepackage[psamsfonts]{amssymb}
\usepackage{amsmath, amsthm, anysize, enumerate, color, url}
\usepackage{graphicx}
\usepackage{epstopdf}
\usepackage{epsfig}
\usepackage{subfigure}
\usepackage[ruled,linesnumbered]{algorithm2e}
\usepackage[sectionbib]{natbib}

\usepackage{threeparttable}

\usepackage[colorlinks, breaklinks,
                   linkcolor=red,
                  citecolor=blue
                  ]{hyperref}

\usepackage{breakurl}

\newtheorem{theorem}{Theorem} 
\newtheorem{corollary}{Corollary}

\newtheorem{lemma}{Lemma} 
\newtheorem{proposition}{Proposition} 

\theoremstyle{definition}
\newtheorem{remark}{Remark}  

\newtheorem{example}{Example}
\newtheorem{condition}{Condition}

\newcommand{\E}{\mathbb{E}}

\newcommand{\R}{\mathbb{R}}

\renewcommand{\P}{\mathbb{P}}

\newcommand{\bs}{\boldsymbol}

\usepackage{setspace}

\begin{document}

\title{A Unified Framework for Inference in Network Models with Degree Heterogeneity and Homophily}

\author{Ting Yan\thanks{Department of Statistics, Central China Normal University, Wuhan, 430079, China.
\texttt{Email:} tingyanty@mail.ccnu.edu.cn.}
\\
Central China Normal University
}
\date{}

\maketitle

\begin{abstract}
The degree heterogeneity and homophily are two typical features in network data.
In this paper, we formulate a general model for undirected networks with these two features
and present the moment estimation for inferring the degree and homophily parameters.
The binary or nonbinary network edges are simultaneously considered.
We establish a unified theoretical framework under which the consistency of the moment estimator holds as the size of networks goes to infinity.
We also derive the asymptotic representation of the moment estimator that can be used to characterize its limiting distribution.
The asymptotic representation of the moment estimator of the homophily parameter contains a bias term.
Two applications are provided to illustrate the theoretical result.
Numerical studies and a real data analysis demonstrate our theoretical findings.


\end{abstract}

\vskip 5pt


\section{Introduction}

Networks/graphs provide a natural way to represent many complex interactive behaviors among a set of actors,
where each node represents an actor and an edge exists between two nodes if the two corresponding actors interact in some
way. The types of interactions could be
friendships between peoples, follow between users in social media such as Twitter, citations between papers, hyperlinks between web pages and so on.
The edges can be directed or undirected, binary (when
each edge is either present or absent) or weighted (when a weight value is recorded).
With the demand of research for a variety of purposes, more and more network data sets have been collected and stored.
At the same time, statistical network analysis have made great process in recent years
and many approaches have been developed; see \cite{Goldenberg2010}, \cite{Fienberg2012}, \cite{Salter-Townshend:2012}, \cite{Advani2018} for some recent reviews.
The book by \cite{Kolaczyk2009} provided a comprehensive coverage of statistical analyses of network data.

One of the most important features of network data is the \emph{degree heterogeneity}, which characterizes the variation in the node degrees.
For example, in the Brightkite network dataset [\cite{Cho:2011}], 
the node degree varies from the minimum value $1$ to the maximum value $1134$ 
in its largest connected subgraph with $56,739$ nodes.
To model the degree heterogeneity, a class of the so-called node-parameter models are proposed, 
in which each node degree is attached to one parameter.
\cite{Holland:Leinhardt:1981} is generally acknowledged as the first one to model the degree variation.
They proposed the $p_1$ model in which
the bi-degrees of nodes and the number of reciprocated dyads form the sufficient statistics for the exponential family distribution on directed graphs.
Other node-parameter models include the Chung-Lu model [\cite{Chung:Lu:2002}] with the expected degrees as the parameters, the $\beta$-model [\cite{Chatterjee:Diaconis:Sly:2011,Blitzstein:Diaconis:2011,Park:Newman:2004,Yan:Xu:2013}],
null models [\cite{Perry:Wolfe:2012}] and maximum entropy models for weighted graphs [\cite{Hillar:Wibisono:2013}].
In these models, the number of parameters increases as the network size grows, so
asymptotic inference is nonstandard.
\cite{Chatterjee:Diaconis:Sly:2011} established the consistency of the maximum likelihood estimator (MLE) in the $\beta$--model
and \cite{Rinaldo2013} derived the necessary and sufficient conditions of its existence. \cite{mukherjee2018} studied the asymptotic properites of some test statistics  for testing sparse signals in the $\beta$-model.
\cite{Hillar:Wibisono:2013} obtained the consistency of the MLE in the maximum entropy models and \cite{Yan:Zhao:Qin:2015}
derived the asymptotically normal distribution of the MLE. \cite{Zhang:Chen:2013} establish a sequential importance sampling method for sampling networks with a given degree sequence.
The degree heterogeneity, directly or indirectly, is also incorporated into other
network models such as the stochastic block model for community detection [\cite{Karrer:Newman:2011, gao2018}], which could give significantly improved fits to network data.

Another important feature, which commonly exists in social and econometric network data, is the \emph{homophily} on individual-level attributes--a phenomenon that individuals tend to form connections with those like themselves [e.g., \cite{McPherson-Lynn2001, kossinets-watts-2006, Sergio-Matthew-2009}].
The individual attributes may be immutable characteristics such as racial and ethnic groups and ages;
it may  also be mutable characteristics such as home address, occupations, levels of
affluence, and personal interests.
The presence of homophily has important implications on the network formation process.
On the one hand, it produces preferential selection--individuals tend more likely to interact with those with similar characteristics.
On the other hand, the existing links create social influence: people may modify their behaviors to bring
them closely into alignment with their neighbors in the network.

The link formation is affected  not only by the homophily effect but also the degree effect.
Neglecting each other might lead to incorrect inference [e.g., \cite{Graham2017}].
To simultaneously model these two features,
\cite{Graham2017} proposed a link surplus model in which a link between two nodes is present
only if the sum of a degree component and a homophily component exceeds a latent random variable
drawn from the logistic distribution.
\cite{Graham2017} derived the consistency and asymptotic normality of the MLE of the homophily parameter.
\cite{Dzemski2019} and \cite{Yan-Jiang-Fienberg-Leng2018} obtained the consistency and asymptotic distribution of
the MLE in the directed link surplus model in which the latent random variables are drawn from the
bivariate normal distribution and the logistic distribution, respectively.
If the focus is only on the homophily parameter, then the conditional method can be used to
eliminate the degree parameters in the case of logistic distribution [\cite{Graham2017,Jochmans2017}].
Another way to address the degree parameters is to treat them as the random effects and
inference is performed by using Bayesian methods [e.g., \cite{Duijn-2004, Krivitsky-20091087, Mele:2017}]. 
In contrast to the random effects method, the joint distribution of the degree heterogeneity and
homophily component is left unrestricted in the fixed effects method.

The contributions of this paper are as follows. First, we formulate a general network model with degree effects and homophily effects
for weighted or unweighted graphs. 
The model here generalizes previous works [e.g., \cite{Graham2017,Dzemski2019,Yan-Jiang-Fienberg-Leng2018}] that only consider binary edges, to  weighted edges.
Second, we establish a unified theoretical framework under which the consistency of the moment estimator in the general network model holds as the number of nodes goes to infinity.
It is notable that the asymptotic results in \cite{Graham2017} are based on the restricted MLE that
restricts the solution of the maximum likelihood problem into a compact set. Our estimator here is left unrestricted.
Furthermore, our result is general, not restricted to a specified distribution.  
Third, we derive the asymptotic representation of the moment estimator that can be used to characterize its limiting distribution.
If the central limit theorem holds for the sum of the observed dyads, then
the moment estimator  converges in distribution to the normal distribution.
The asymptotic representation of the estimator of the homophily parameter contains a bias term.
Valid inference requires bias-correction.
Finally, the unified theoretical framework is illustrated by two applications as well as numerical simulations.
A real data analysis is also provided.

For the rest of the paper, we proceed as follows. In Section \ref{section:model}, we introduce the model.
In Section \ref{section:estimation}, we present the estimation method.
In Section \ref{section:asymptotic}, we present the consistency and asymptotic normality of the moment estimator.
In Section \ref{section:app}, we illustrate the main result by an application.
We give the summary and further discussion in Section \ref{section:sd}.
The proofs of the main results are regelated into Section \ref{section:appendix}.
The proofs of some propositions and lemmas are given in the supplementary material.

\input{main}

\setlength{\itemsep}{-1.5pt}
\setlength{\bibsep}{0ex}
\bibliography{reference3}
\bibliographystyle{apa}

\end{document}

%% file: main.tex
\section{Model}
\label{section:model}

Let $G_n$ be an undirected graph on $n\geq 2$ nodes labeled by ``$1, \ldots, n$".
Let $A=(a_{ij})_{n\times n}$ be the adjacency matrix of $G_n$, where
$a_{ij}$ is the weight of the edge between nodes $i$ and $j$.
We do not consider self-loops here, i.e., $a_{ii}=0$.
The graph $G_n$ may be weighted or unweighted. If the edge weight $a_{ij}$ is
an indicator (present or absent), then $G_n$ is unweighted (or called a simple graph).
If $a_{ij}$ takes values from a set of positive integers (e.g., the number of papers collaborated by authors $i$ and $j$ in coauthor networks),
then the graph $G_n$ is weighted. Moreover, $a_{ij}$ could be continuous (e.g., the call time between two peoples).
Let $d_i= \sum_{j \neq i} a_{ij}$ be the degree of node $i$
and $d=(d_1, \ldots, d_n)^\top$ be the degree sequence of the graph $G_n$.
We also observe a vector $z_{ij}$, the covariate information attached to the edge between nodes $i$ and $j$.
The covariate $z_{ij}$ can be formed according to the similarity or dissimilarity between
 node attributes $x_i$ and $x_j$ for nodes $i$ and $j$. Specifically, $z_{ij}$ can be represented through a symmetric function $g(\cdot, \cdot)$ with $z_i$ and $z_j$ as its arguments. As an example
if $x_{i1}$ and $x_{i2}$ are location coordinates, then $z_{ij} =[(x_{i1}- x_{j1})^2 + (x_{i2} - x_{j2})^2]^{1/2}$ denotes the Euclidean distance
between $i$ and $j$.

We mainly focus on network models with two typical network features: the degree heterogeneity and homophily.
The first is measured by a set of unobserved degree parameters $\{\beta_i\}_{i=1}^n$ and the second by the regression coefficient $\gamma$ of the pairwise covariates.
Following \cite{Graham2017}, we assume that all edges are independent.
We assume that the probability density function of the edge variable $a_{ij}$ conditional on the unobserved degree effects and
observed covariates has the following form:
\begin{equation}
\label{model}
a_{ij}=a|z_{ij}, \gamma, \beta_i, \beta_j \sim  f(a|z_{ij}, \gamma, \beta_i, \beta_j),
\end{equation}
where $f$ is a known probability density function, $\beta_i$ is the degree parameter of node $i$ and $\gamma$ is a $p$-dimensional coefficient for the covariate $z_{ij}$.
Throughout the paper, we assume that $p$ is fixed.
The parameter $\beta_i$ is the intrinsic individual effect that reflects the node heterogeneity to participate in network connection.
The common parameter $\gamma$ is exogenous, measuring the homophily effect.
If $f(\cdot)$ is an increasing function of $\beta_i$,
then those nodes having relatively large degree parameters will have more links than those nodes with low degree parameters when neglecting the homophily effect.
A larger homophily component $z_{ij}^\top \gamma$ means a larger homophily effect.
Two running examples for illustrating the model are given below.


\begin{example}(Binary weight)
\label{example-a}
Let $a_{ij}$ be the binary weight of edge $(i,j)$, i.e., $a_{ij}\in \{0,1\}$, and
$F$ be a cumulative distribution function.
The probability of $a_{ij}$ is
\[
\P( a_{ij}= a ) = [F( \beta_i+\beta_j + z_{ij}^\top \gamma )]^{a} (1-F(\beta_i+\beta_j+z_{ij}^\top \gamma ) )^{1-a},~~a=0, 1.
\]
Two common examples for $F(\cdot)$ are the logistic distribution: $F(x)=e^x(1+e^x)^{-1}$ [e.g., \cite{Graham2017}] and probit distribution: $F(x)=\Phi(x)$.
Here, $\Phi(x)$ is the cumulative distribution function for the standard normal random variable.
\end{example}

\begin{example}(Infinite discrete weight)
\label{example-b}
Let $a_{ij}\in \{0, 1, \ldots\}$.  The probability of the edge weight is assumed to be distributed by a Poisson distribution with mean $\lambda= \exp( \beta_i+\beta_j + z_{ij}^\top \gamma)$, i.e.,
\[
\log \P(a_{ij}=a) = a(\beta_i+\beta_j + Z_{ij}^\top \gamma) - \exp (\beta_0 + \beta_i+\beta_j + Z_{ij}^\top \gamma) - \log a!.
\]
\end{example}

To establish a unified theoretical result, we need to make a basic model assumption.

\emph{Model assumption}. We assume that the degree parameters enter the marginal probability density function $f(\cdot)$
additively through $\beta_i+\beta_j$. Further, the additive structure also applies to the homophily component.
That is,
\[
a_{ij}|z_{ij}, \beta, \gamma \sim f(a| z_{ij}^\top \gamma + \beta_i + \beta_j ).
\]

The dependence of the distribution $f(\cdot)$ on the
parameters is through an index $z_{ij}^\top \gamma + \beta_i + \beta_j$ as given in the above examples.
This is referred to as single index models in econometrics literature.
We focus on these additive models for computational tractability. However, the method developed in this paper can be adapted to
the non-additive structure for both effects.

\section{Estimation}
\label{section:estimation}
Write $\mu(\cdot)$ as the expectation on the distribution $f(\cdot)$.
Define
\begin{equation}\label{definition-pi}
\pi_{ij}:=z_{ij}^\top \gamma + \beta_i + \beta_j.
\end{equation}
Since the dependence of the expectation of $a_{ij}$ on parameters is only
through $\pi_{ij}$,
we can write $\mu(\pi_{ij})$ as the expectation of $a_{ij}$.
When we emphasize the arguments $\beta$ and $\gamma$ in $\mu(\cdot)$, we write $\mu_{ij}(\beta, \gamma)$ instead of $\mu(\pi_{ij})$.
To estimate the parameters, we use the moment estimation.
The moment equations are as follows:
\renewcommand{\arraystretch}{1.2}
\begin{equation}\label{eq:moment}
\begin{array}{rcl}
d_i & = & \sum_{j\neq i} \mu_{ij}(\beta, \gamma),~~ i=1, \ldots, n, \\
\sum_{i=1}^n \sum_{j=1, j<i}^n a_{ij}z_{ij} & = & \sum_{i=1}^n \sum_{j=1, j<i}^n   z_{ij} \mu_{ij}(\beta, \gamma).
\end{array}
\end{equation}
The solution to the above equations is the moment estimator denoted by $(\widehat{\beta}, \widehat{\gamma})$.
Let
\[
\varphi(\beta, \gamma)= (\sum_{j\neq 1} \mu_{1j}(\beta, \gamma), \ldots, \sum_{j\neq n} \mu_{nj}(\beta, \gamma) ,  \sum_{i=1}^n \sum_{j=1, j<i}^n     \mu_{ij}(\beta, \gamma)z_{ij}^\top )^\top.
\]
If the inverse function $\varphi^{-1}$ exists, then the moment estimator of $(\beta, \gamma)$ exists and is unique, i.e.,  $(\widehat{\beta}, \widehat{\gamma})=\varphi^{-1}(d, \sum_{i<j} a_{ij}z_{ij})$.
When $\varphi^{-1}$ does not exist (i.e.,  $\varphi$ is not one-to-one), any solution $(\widehat{\beta}, \widehat{\gamma})$ of equation \eqref{eq:moment}
is a moment estimator of $(\beta, \gamma)$. In some cases, a moment estimator may not exist. Under some regularity conditions,
the moment estimator exists with a large probability. The details are given in next section.

Now we discuss some computational issues.
When the number of nodes $n$ is small and $f$ is the binomial, Probit, or Poisson probability function or Gamma density function,
we can simply use the package ``glm" in the R language to solve \eqref{eq:moment}.
For relatively large $n$, it might not have large enough memory to store the design matrix for $\beta$ required by the R package ``glm".
In this case,
we recommend the use of a two-step iterative algorithm by alternating between solving the first equation in  \eqref{eq:moment} via the fixed point method in \cite{Chatterjee:Diaconis:Sly:2011}
or other numerical methods
and solving the second equation in \eqref{eq:moment} via an iteratively reweighted least squares method
for generalized linear models [\cite{McCullagh-Nelder-1989}].

\section{Asymptotic properties}
\label{section:asymptotic}
In this section, we present the consistency and asymptotic representation of the moment estimator.
We first introduce some notations. For a subset $C\subset \R^n$, let $C^0$ and $\overline{C}$ denote the interior and closure of $C$, respectively.
For a vector $x=(x_1, \ldots, x_n)^\top\in R^n$, denote by $\|x\|$ for a general norm on vectors with the special cases
$\|x\|_\infty = \max_{1\le i\le n} |x_i|$ and $\|x\|_1=\sum_i |x_i|$ for the $\ell_\infty$- and $\ell_1$-norm of $x$ respectively.
When $n$ is fixed, all norms on vectors are equivalent. Let $B(x, \epsilon)=\{y: \| x-y\|_\infty \le \epsilon\}$ be an $\epsilon$-neighborhood of $x$.
For an $n\times n$ matrix $J=(J_{i,j})$, let $\|J\|_\infty$ denote the matrix norm induced by the $\ell_\infty$-norm on vectors in $\R^n$, i.e.,
\[
\|J\|_\infty = \max_{x\neq 0} \frac{ \|Jx\|_\infty }{\|x\|_\infty}
=\max_{1\le i\le n}\sum_{j=1}^n |J_{i,j}|,
\]
and $\|J\|$ be a general matrix norm.
Define the matrix maximum norm: $\|J\|_{\max}=\max_{i,j}|J_{ij}|$.
We use the superscript ``*" to denote the true parameter under which the data are generated.
When there is no ambiguity, we omit the super script ``*".
Define
\begin{equation}\label{definition-kappa}
\kappa_n := \max_{i,j} \| z_{ij} \|_\infty.
\end{equation}
When causing no confusion, we will simply write $\mu_{ij}$ stead of $\mu_{ij}(\beta, \gamma)$ for shorthand.
The notation $\sum_{j<i}$  is a shorthand for $\sum_{i=1}^n \sum_{j=1, j<i}^n$.

Throughout the paper, we assume that $\mu(\cdot)$ is a continuous function with the third derivative.
Recall that $\pi_{ij}=\beta_i+\beta_j+z_{ij}^\top \gamma$ defined at \eqref{definition-pi}.
Write $\mu^\prime$, $\mu^{\prime\prime}$ and $\mu^{\prime\prime\prime}$ as the first, second and third derivative of $\mu(\pi)$ on $\pi$, respectively.
Let $\epsilon_{n1}$ and $\epsilon_{n2}$ be two small positive numbers.
When $\beta \in B(\beta^*, \epsilon_{n1}), \gamma\in B(\gamma^*, \epsilon_{n2})$, we assume that there are four positive numbers $b_{n0}, b_{n1}, b_{n2}, b_{n3}$ such that
\begin{subequations}
\begin{gather}
\label{ineq-mu-keya}
b_{n0}\le \min_{i,j} |\mu^\prime(\pi_{ij})| \le \max_{i,j}|\mu^\prime(\pi_{ij})|\le b_{n1},  \\
\label{ineq-mu-keyb}
\max_{i,j}|\mu^{\prime\prime}(\pi_{ij})| \le b_{n2}, \\
\label{ineq-mu-keyc}
\max_{i,j}|\mu^{\prime\prime\prime}(\pi_{ij})| \le b_{n3}.
\end{gather}
\end{subequations}
Under the above inequalities, the following holds:
\begin{eqnarray}
\label{ineq-mu-beta}
\max_{i,j}\sup_{\beta \in B(\beta^*, \epsilon_{n1})} |\mu_{ij}(\beta, \gamma^*) - \mu_{ij}(\beta^*, \gamma^*) | \le 2b_{n1} \| \beta - \beta^* \|_\infty ,\\
\label{ineq-mu-gamma}
\max_{i,j}\sup_{\gamma \in B(\gamma^*, \epsilon_{n1})} |\mu_{ij}(\beta^*, \gamma) - \mu_{ij}(\beta^*, \gamma^*) | \le b_{n1}\kappa_n \|\gamma - \gamma^*\|_1.
\end{eqnarray}

\subsection{Consistency}
To deduce the conditions of the consistency for the moment estimator, let us first define a system of functions based on the moment equations.
Define
\begin{equation}\label{eqn:def:F}
 F_i(\beta, \gamma)= d_i - \sum\limits_{j=1, j\neq i}^n \mu_{ij}(\beta, \gamma),~~i=1, \ldots, n,
\end{equation}
and $F(\beta, \gamma)=(F_1(\beta, \gamma), \ldots, F_n(\beta, \gamma))^\top$.
Further, we define $F_{i,\gamma}(\beta)$ as the value of $F_i(\beta, \gamma)$ for an arbitrarily fixed $\gamma$ 
and $F_\gamma(\beta)=(F_{1,\gamma}(\beta), \ldots, F_{n,\gamma}(\beta))^\top$.
Let $\widehat{\beta}_\gamma$ be a solution to $F_\gamma(\beta)=0$.
Correspondingly, we define two functions for exploring the asymptotic behaviors of the estimator of the homophily parameter:
\begin{eqnarray}
\label{definition-Q}
Q(\beta, \gamma)= \sum_{j<i} z_{ij} (a_{ij}  -   \mu_{ij}(\beta, \gamma) ), \\
\label{definition-Qc}
Q_c(\gamma)= \sum_{j<i} z_{ij} (a_{ij}  -    \mu_{ij}(\widehat{\beta}_\gamma, \gamma) ).
\end{eqnarray}
$Q_c(\gamma)$ could be viewed as the concentrated or profile function of the moment function $Q(\beta, \gamma)$ in which the degree parameter $\beta$ is profiled out.
It is clear that
\begin{equation}\label{equation:FQ}
F(\widehat{\beta}, \widehat{\gamma})=0,~~F(\widehat{\beta}_\gamma, \gamma)=F_\gamma(\widehat{\beta}_\gamma)=0,~~Q(\widehat{\beta}, \widehat{\gamma})=0,~~Q_c(\widehat{\gamma})=0.
\end{equation}

If the moment estimator ($\widehat{\beta}, \widehat{\gamma})$ is consistent, then it is natural to require that the norm $\| F(\beta, \gamma) \|_\infty$ evaluated at the true parameters $\beta^*$ and $\gamma^*$ is small.
This leads to our first condition.

\begin{condition}\label{condition-diff-a}
For $F(\beta, \gamma)$ defined at \eqref{eqn:def:F}, we require that
\[
\| F(\beta^*, \gamma^*) \|_\infty = O_p( h_{n1}\sqrt{n\log n} ),
\]
where $h_{n1}$ is a scalar factor that may depend on the ranges of $\beta^*$ and $\gamma^*$.
\end{condition}

Condition \ref{condition-diff-a} requires that $F_i(\beta^*, \gamma^*)$ is in the order of $(n\log n)^{1/2}$. It can be verified as follows.
If the sequence $\{ a_{ij} \}_{j=1}^n$ is independent for any fixed $i$,
then Condition \ref{condition-diff-a} holds in the light of \citeauthor{Hoeffding:1963}'s inequality for bounded random variables or concentration inequality for sub-exponential random variables
[e.g., Corollary 5.17 in \citeauthor{vershynin_2012} (\citeyear{vershynin_2012})].
These probability inequalities depend on the values of parameters that leads to the additional factor $h_{n1}$.
More specifically, $h_{n1}$ depends on $\|\beta^*\|_\infty$ and $\|\gamma^*\|_\infty$.
If $\|\beta^*\|_\infty$ and $\|\gamma^*\|_\infty$ are bounded by a constant, then $h_{n1}$ is also a constant, regardless of $n$.

Let $F(x): \R^n \to \R^n$ be a function vector on $x\in\R^n$. We say that a Jacobian matrix $F^\prime(x)$ with $x\in \R^n$ is Lipschitz continuous on a convex set $D\subset\R^n$ if
for any $x,y\in D$, there exists a constant $\lambda>0$ such that
for any vector $v\in \R^n$ the inequality
\begin{equation*}
\| [F^\prime (x)] v - [F^\prime (y)] v \|_\infty \le \lambda \| x - y \|_\infty \|v\|_\infty
\end{equation*}
holds.
The proposition below shows that $F_{\gamma}(\beta)$ is Lipschitz continuous, whose proof is given in the supplementary material.

\begin{proposition}\label{condition:lipschitz-c}
Let $D=B(\beta^*, \epsilon_{n1}) (\subset \R^{n})$ be an open convex set containing the true point $\beta^*$.
For $\gamma\in B(\gamma^*, \epsilon_{n2})$, if inequality \eqref{ineq-mu-keyc} holds, then
the Jacobian matrix $F'_\gamma( x )$ of $F_\gamma(x)$ on $x$ is Lipschitz continuous on $D$ with the Lipschitz coefficient  $4b_{n2}(n-1)$.
\end{proposition}

The Lipschitz continuous property of $F_\gamma^\prime$ is one of the conditions to guarantee  the consistency of the moment estimator.
The Jacobian matrix $F'_\gamma( \beta )$ of $F_\gamma(\beta)$ on the parameter $\beta$ has a special structure that can be characterized in the form of a matrix class.
Given $m, M>0$, we say an $n\times n$ matrix $V=(v_{ij})$ belongs to the matrix class $\mathcal{L}_{n}(m, M)$ if
$V$ is a diagonally balanced matrix with positive elements bounded by $m$ and $M$, i.e.,
\begin{equation}\label{eq1}
\begin{array}{l}
v_{ii}=\sum_{j=1, j\neq i}^{n} v_{ij}, ~~i=1,\ldots, n, \\
m\le v_{ij} \le M, ~~ i,j=1,\ldots,n; i\neq j.
\end{array}
\end{equation}
It can be easily checked that  $F'_\gamma(\beta)$ or $-F'_\gamma(\beta)$ belongs to this matrix class.
We will obtain the consistency of the estimator $\hat{\beta}_\gamma$ through
the convergence rate of the Newton iterative method, which depends on the inverse of  $F'_\gamma( \beta )$.
We describe the characterization of the Jacobian matrix $F'_\gamma( \beta )$ in terms of the following proposition.

\begin{proposition}\label{condition:matrixclass-d}
Assume that $\beta \in B(\beta^*, \epsilon_{n1})$ and $\gamma \in B(\gamma^*, \epsilon_{n2})$. If inequality \eqref{ineq-mu-keya} holds,
then $F'_{\gamma}(\beta) \in\mathcal{ L}_n(b_{n0}, b_{n1})$ or $-F'_{\gamma}(\beta) \in \mathcal{L}_n(b_{n0}, b_{n1})$.
\end{proposition}

The following lemma characterizes the upper bound of the error between $\widehat{\beta}_\gamma$ and $\beta^*$.

\begin{lemma}\label{lemma-a}
Let $\epsilon_{n1}$ be a positive number and $\epsilon_{n2}=(\log n/ n)^{1/2}$.
Under Condition \ref{condition-diff-a}, if inequalities \eqref{ineq-mu-keya}, \eqref{ineq-mu-keyb} and \eqref{ineq-mu-keyc}  hold and
\begin{equation}\label{equation:lemma-a}
\frac{b_{n1}^4b_{n2} (h_{n1}+b_{n1}\kappa_n)(\log n)^{1/2}}{n^{1/2}b_{n0}^6}    = o(1),
\end{equation}
then with probability approaching one, $\widehat{\beta}_\gamma$ exists and satisfies
\[
\| \widehat{\beta}_\gamma - \beta^* \|_\infty = O_p\left( \frac{ b_{n1}^2(h_{n1}+ b_{n1}\kappa_n)}{b_{n0}^3}\sqrt{\frac{\log n}{n}} \right) = o_p(1).
\]
\end{lemma}

To show the consistency of $\widehat{\gamma}$, similar to Condition \ref{condition-diff-a},
we need the following condition.
\begin{condition}\label{condition-Q}
$ \|Q(\beta^*, \gamma^*) \| = O_p( h_{n2} n^{3/2}\log n )$, where $h_{n2}$ is a scalar factor.
\end{condition}

The above condition is mild. If $a_{ij}$'s ($i<j$) are independent, then the upper bound of $ \|Q(\beta^*, \gamma^*) \|$ is in the magnitude of $n$.
Examples are given in next section.
Similar to Proposition \ref{condition:lipschitz-c}, we have the following proposition, whose proof is given in the supplementary material.

\begin{proposition}\label{proposition-Q-Lip}
Let $D=B(\gamma^*, \epsilon_{n2}) (\subset \R^{p})$ be an open convex set containing the true point $\gamma^*$.
Assume that  \eqref{ineq-mu-keya}, \eqref{ineq-mu-keyb}, \eqref{ineq-mu-keyc} and \eqref{equation:lemma-a}  hold.
If $\| F(\beta^*, \gamma^*) \|_\infty = O( h_{n1}(n\log n)^{1/2} )$, then
$ Q_c(\gamma)$ is Lipschitz continuous on $D$ with the Lipschitz coefficient  $n^2 \kappa_n^4 b_{n1}^{11} b_{n2} b_{n0}^{-9}$. 
\end{proposition}

The asymptotic behavior of $\widehat{\gamma}$ crucially depends on the Jacobian matrix $Q_c^\prime(\gamma)$.
The expression for the derivative of $Q_c(\gamma)$ on $\gamma$ is
\begin{eqnarray}\label{equation:Qc-derivative}
\frac{ \partial Q_c(\gamma) }{ \partial \gamma^\top } & = &
\frac{ \partial Q }{ \partial \gamma^\top} \bigg |_{\beta=\widehat{\beta}_\gamma,\gamma=\gamma}
\\
\nonumber
&& - \frac{ \partial Q }{\partial \beta^\top}\bigg |_{\beta=\widehat{\beta}_\gamma,\gamma=\gamma}
 \left[\frac{\partial F}{\partial \beta^\top} \bigg |_{\beta=\widehat{\beta}_\gamma,\gamma=\gamma} \right]^{-1}
\frac{\partial F}{\partial \gamma^\top}\bigg |_{\beta=\widehat{\beta}_\gamma,\gamma=\gamma},
\end{eqnarray}
where $Q=Q(\beta, \gamma)$ and $F=F(\beta, \gamma)$.
Since $\widehat{\beta}_\gamma$ does not have a closed form, conditions that are directly imposed on $Q_c^\prime(\gamma)$ are not easily checked.
To derive feasible conditions, we define
\begin{equation}
\label{definition-H}
H(\beta, \gamma) = \frac{ \partial Q(\beta, \gamma) }{ \partial \gamma} - \frac{ \partial Q(\beta, \gamma) }{\partial \beta} \left[ \frac{\partial F(\beta, \gamma)}{\partial \beta} \right]^{-1}
\frac{\partial F(\beta, \gamma)}{\partial \gamma},
\end{equation}
which is a general form of $ \partial Q_c(\gamma) / \partial \gamma$.
Note that the dimension of $H(\beta, \gamma)$ is fixed. All matrix norms on $H(\beta, \gamma)$ are equivalent.
The next condition bounds the matrix norm $\|Q_c(\gamma)\|$ of $Q_c(\gamma)$.

\begin{condition}\label{condition-H}
For $\beta\in B(\beta^*, \epsilon_n)$, it is required that $\| H^{-1}(\beta, \gamma^*) \| = O( h_{n3}/n^2  )$, where $h_{n3}$ is a scalar factor.
\end{condition}

When $\beta\in B(\beta^*, \epsilon_n)$, we have the equation:
\begin{equation}\label{equation-H-appro}
\frac{1}{n^2} H(\beta, \gamma^*) = \frac{1}{n^2} H(\beta^*, \gamma^*) + o(1),
\end{equation}
whose proof is given in the supplementary material.
Now we formally state the consistency result.

\begin{theorem}\label{Theorem:con}
Assume that \eqref{ineq-mu-keya}, \eqref{ineq-mu-keyb} and \eqref{ineq-mu-keyc} hold.
Under Conditions \ref{condition-diff-a}--\ref{condition-H}, if equation \eqref{equation:lemma-a} and the following equation hold:
\begin{equation}\label{eq:consistency}
\eta_n h_{n3}^2  \kappa_n^4 b_{n1}^{11} b_{n2} b_{n0}^{-9} n^{-1/2}(\log n)^{1/2} = o(1),
\end{equation}
where $\eta_n=h_{n2} +\kappa_n b_{n1}^3 (h_{n1}+b_{n1}\kappa_n)b_{n0}^{-3}$,
then the moment estimator $\widehat{\gamma}$ exists with probability approaching one and is consistent in the sense that
\begin{equation}\label{Newton-convergence-rate}
\| \widehat{\gamma} - \gamma^{*} \|_\infty = O_p\left( \eta_n h_{n3} \sqrt{\frac{\log n}{n}} \right)=o_p(1)
\end{equation}
and
\[
\| \widehat{\beta} - \beta^* \|_\infty = O_p\left( \frac{ b_{n1}^2(h_{n1}+b_{n1}\kappa_n)}{b_{n0}^3}\sqrt{\frac{\log n}{n}} \right)=o_p(1).
\]
\end{theorem}

\subsection{Asymptotic representation}
Let $T_{ij}$ be a vector of length $n$ with $i$th and $j$th elements ones and other elements zeros. Define
\[
s_{\beta_{ij}}(\beta, \gamma)= (a_{ij} - \mu_{ij}(\beta, \gamma))T_{ij}, ~~s_{\gamma_{ij}}(\beta, \gamma)= z_{ij}(a_{ij} - \mu_{ij}(\beta, \gamma)).
\]
Let
\[
V(\beta, \gamma)=\frac{ \partial F(\beta, \gamma) }{ \partial \beta^\top }, ~~
V_{\gamma\beta}(\beta, \gamma) = \frac{ \partial Q(\beta, \gamma) }{ \partial \beta^\top}.
\]
Then we define
\[
\tilde{s}_{\gamma_{ij}} (\beta, \gamma)=s_{\gamma_{ij}}(\beta, \gamma)-V_{\gamma\beta}(\beta, \gamma)[V(\beta, \gamma)]^{-1}s_{\beta_{ij}}(\beta, \gamma).
\]
Let $N=n(n-1)$ and
\[
\bar{H}=\lim_{n\to\infty} \frac{1}{N} H( \beta^*, \gamma^*),
\]
where $H(\beta, \gamma)$ is defined at \eqref{definition-H}.
We assume that the above limit exists.
The asymptotic representation of $\widehat{\gamma}$ is stated below.

\begin{theorem}
\label{theorem-central-b}
Assume that the conditions in Theorem \ref{Theorem:con} hold.
If
\[
\frac{b_{n3}\kappa_n b_{n1}^6(h_{n1} + b_{n1}\kappa_n )^3(\log n)^{3/2}}{n^{1/2}b_{n0}^9}=o(1),
\]
then we have
\[
\sqrt{N}(\widehat{\gamma}- \gamma^*) = \bar{H}^{-1} B_* + \bar{H}^{-1} \frac{1}{\sqrt{N}} \sum_{j< i}
\tilde{s}_{\gamma_{ij}} (\beta^*, \gamma^*) + o_p(1),
\]
where
\begin{equation}\label{defintion-Bias}
B_*=\lim_{n\to\infty} \frac{1}{2\sqrt{N}} \sum_{i=1}^n \frac{  \sum_{j\neq i} z_{ij} \mu_{ij}^{\prime\prime}(\pi^*_{ij})  }
{  \sum_{j\neq i} \mu_{ij}^\prime(\pi^*_{ij})  },
\end{equation}
and $\pi^*_{ij}=\beta_i^* + \beta_j^* + z_{ij}^\top \gamma^*$.
\end{theorem}

\begin{remark}
The asymptotic expansion of $\widehat{\gamma}$ contains a bias term $B_*$. 
If the parameter vector $\beta$ and all homophily components $z_{ij}^\top \gamma$'s are bounded, then
$\|\bar{H}\| = O(1)$ and $\|B_*\|_\infty=O(1)$.
It follows that $\widehat{\gamma}$ has a convergence rate at around $n^{-1}$.
Since $\widehat{\gamma}$ is not centered at the true parameter value, the confidence intervals
and the p-values of hypothesis testing constructed from $\widehat{\gamma}$ cannot achieve the nominal level without bias-correction. 
This is referred to as the well-known incidental parameter problem in econometrics literature [\cite{Neyman:Scott:1948, FVW2016, Dzemski2019}].
The produced bias is due to the appearance of additional parameters.
Here, we propose to use the analytical bias correction formula:
$\widehat{\gamma}_{bc} = \widehat{\gamma}- N^{-1/2}H^{-1}(\widehat{\beta},\widehat{\gamma}) \hat{B}$,
where $\widehat{B}$ is the estimate of $B_*$ by replacing
$\beta^*$ and $\gamma^*$ in their expressions with their estimators $\widehat{\beta}$ and
$\widehat{\gamma}$, respectively.
\end{remark}

\begin{remark}
If $N^{-1/2} \sum_{j< i}
\tilde{s}_{\gamma_{ij}} (\beta^*, \gamma^*)$ asymptotically follows a multivariate normal distribution,
then $\sqrt{N}(\widehat{\gamma}- \gamma^*)$ converges in distribution to the normal distribution.
\end{remark}

\begin{theorem}\label{Theorem-central-a}
Let $S=\mathrm{diag}(1/[V(\beta^*, \gamma^*)]_{11}, \ldots, 1/[V(\beta^*, \gamma^*)]_{nn})$
 and $R=\mathrm{Cov}( d - \E d )$.
Assume that the conditions in Theorem \ref{Theorem:con} hold.
If
\[
\frac{b_{n1}^2(\varphi_{n1}^2+\varphi_{n2}^2\kappa_n^2)b_{n2}}{b_{n0}^3} \frac{\log n}{n^{1/2}}=o(1),
\]
and
\[
\frac{b_{n1}^2\kappa_n \| \widehat{\gamma}-\gamma^* \|_1 }{b_{n0}^3} = o_p(n^{-1/2}), ~~ \max_i \frac{b_{n1}^4}{n^4 b_{n0}^6} \sum_{i,j}|R_{ij}|  = o( \frac{1}{n} ).
\]
then for any fixed $i$,
\begin{equation*}
\widehat{\beta}_i- \beta_i=  v_{ii}^{-1}(d_i - \E d_i)  + o_p( n^{-1/2} ),
\end{equation*}
where
\[
\varphi_{n1}=b_{n1}^2b_{n0}^{-3}  (h_{n1}+b_{n1}\kappa_n), ~~~ \varphi_{n2}= h_{n3}[h_{n2} + \kappa_n b_{n1}^3 (h_{n1}+ b_{n1}\kappa_n)b_{n0}^{-3}].
\]
\end{theorem}

\begin{remark}
We make a remark about the condition $b_{n1}^2b_{n0}^{-3}\kappa_n \| \widehat{\gamma}-\gamma^* \|_1  = o_p(n^{-1/2})$.
According to the asymptotic expansion of $\widehat{\gamma}$ in Theorem \ref{theorem-central-b}, if $\frac{1}{\sqrt{N}} \sum_{j< i}
\tilde{s}_{\gamma_{ij}} (\beta^*, \gamma^*)$ converges in distribution to the normal distribution, then
$\| \widehat{\gamma}-\gamma^* \|_\infty$ is in the magnitude of $n^{-1}$ with probability approaching one.
So this condition is mild and generally holds.
\end{remark}

\begin{remark}
We discuss the condition $\max_i(WBW^\top)_{ii} = o( n^{-1} )$. If $V=\mathrm{Cov}( d - \E d )$,
then $WBW^\top = V^{-1} -S -S (I - VS)$, where $I$ is the identify matrix of order $n$. 
It is easy to verify that $\|WVW^\top\|_{\max}= O(n^{-2}b_{n1}^2b_{n0}^{-3})$. In this case, $\max_i(WBW^\top)_{ii} = o( n^{-1} )$.
If all random edges $\{a_{ij}\}_{j<i}$ are independent and their distributions belong to the exponential family, then $V=\mathrm{Cov}( d - \E d )$.
\end{remark}

\begin{remark}\label{remark:asymptotic}
If for any fixed $k$, the vector $(d_1 - \E d_1, \ldots, d_k - \E d_k)$  is asymptotically multivariate normal distribution with
mean $0$ and covariance matrix $\Sigma_{kk}$, then the vector
\[
(S_{kk}^{-1}\Sigma_{kk}S_{kk}^{-1})^{-1/2}(\widehat{\beta}_1- \beta_1, \ldots, \widehat{\beta}_k - \beta_k)^\top
\]
converges in distribution to the standard normal distribution by  Theorem \ref{Theorem-central-a}, where
$S_{kk}$ is the upper left $k \times k$ submatrix of $S$.
In the case of edge independence that $\{a_{ij}\}_{j=1}^n$ is an independent random variable sequence for any fixed $i$, the claim that
$v_{ii}^{-1/2}(d_i - \E d_i)$ converges in distribution to the standard normal distribution,  can be checked by various kinds of classical conditions for
the central limit theorem such as Lyapunov's condition [\cite{Billingsley:1995}, page 362] and \citeauthor{Lindeberg1922}'s (\citeyear{Lindeberg1922}) condition.
\end{remark}

\section{Applications}
\label{section:app}
In this section, we illustrate the theoretical result by two applications,
in which we consider the logistic model and Poisson model for $f(\cdot)$, respectively.

\subsection{The logistic model}
We consider the binary weight for each edge, i.e., $a_{ij}\in \{0, 1\}$.
Following \cite{Graham2017}, we assume that all dyads $(a_{ij}, a_{ji})$'s are independent.
Under this assumption, the maximum likelihood equations are identical to the moment equations in \eqref{eq:moment}.
The aim of this application is to relax the assumption that the MLE is restricted into a compact set made by \cite{Graham2017}.
The model is
\[
\P(a_{ij}=1) = \frac{ e^{z_{ij}^\top \gamma + \beta_i + \beta_j }}{ 1 + e^{z_{ij}^\top \gamma + \beta_i + \beta_j } }.
\]
The moment equations are
\begin{equation}\label{eq:likelihood-binary}
\begin{array}{c}
d_i  =  \sum_{j\neq i} \frac{ e^{z_{ij}^\top \gamma + \beta_i + \beta_j }}{ 1 + e^{z_{ij}^\top \gamma + \beta_i + \beta_j } } ,~~~i=1,\ldots, n, \\
\sum_{j<i} a_{ij}z_{ij} = \sum_{j<i}  \frac{ z_{ij}e^{z_{ij}^\top \gamma + \beta_i + \beta_j }}{ 1 + e^{z_{ij}^\top \gamma + \beta_i + \beta_j } }.
\end{array}
\end{equation}
In the this case, $\mu(x)=e^x/(1+e^x)$. It can be shown that
\[
\mu^\prime(x) = \frac{e^x}{ (1+e^x)^2 },~~  \mu^{\prime\prime}(x) = \frac{e^x(1-e^x)}{ (1+e^x)^3 },~~ \mu^{\prime\prime\prime}(x) = \frac{e^x(1-4e^x+e^{2x})}{ (1+e^x)^4 }.
\]
It is easily checked that
\[
|\mu^\prime(x)| \le \frac{1}{4}, ~~ |\mu^{\prime\prime}(x)| \le \frac{1}{4},~~ |\mu^{\prime\prime\prime}(x)| \le \frac{1}{4},
\]
where the last two inequalities are due to
\[
|\mu^{\prime\prime}(x)| \le \frac{e^x}{ (1+e^x)^2 } \times \left|\frac{(1-e^x)}{ (1+e^x) }\right|,
\]
and
\[
|\mu^{\prime\prime\prime}(x)| =
\frac{e^x}{ (1+e^x)^2 } \times \left| \left[ \frac{(1-e^x)^2}{ (1+e^x)^2 } - \frac{2e^x}{ (1+e^x)^2 }  \right]\right|,
\]
respectively.
So $b_{n1}=b_{n2}=b_{n3}=1/4$ in inequalities \eqref{ineq-mu-keya}, \eqref{ineq-mu-keyb} and \eqref{ineq-mu-keyc},
respectively.
Since $f(x)=e^x(1+e^x)^{-2}$ is a decreasing function of $x$ when $x\ge 0$ and an increasing function of $x$ when $x\le 0$, we have that when
$\beta \in B(\beta^*, \epsilon_{n1})$ and  $\gamma\in B(\gamma^*, \epsilon_{n2})$,
\[
b_{n0} = \min_{ i,j} \frac{e^{\pi_{ij}}}{ (1+e^{\pi_{ij}})^2 } \ge  \frac{ e^{ 2\|\beta^*\|_\infty + \|\gamma^*\|_1 \kappa_n + 2\epsilon_{n1}+p\epsilon_{n2} } }{ (1 + e^{ 2\|\beta^*\|_\infty + \|\gamma^*\|_1 \kappa_n + 2\epsilon_{n1}+p\epsilon_{n2} })^2 }.
\]
Note that $d_i$ is a sum of $n-1$ independent Bernoulli random variables. By \citeauthor{Hoeffding:1963}'s \citeyearpar{Hoeffding:1963} inequality, we have
\begin{eqnarray*}
&&P(|d_i - \E(d_i)|\ge \sqrt{(n-1)\log (n-1) }) \\
&\le & 2\exp \left( -\frac{2\left(\sqrt{(n-1)\log (n-1) }\right)^2 }{(n-1)} \right) = \frac{2}{(n-1)^2},
\end{eqnarray*}
such that
\[
 P(\max_i|d_i - \E(d_i)|\ge x) \le \sum_i P(|d_i - \E(d_i)|\ge x) = O(\frac{1}{n}).
\]
Therefore, we have
\[
\max_i |d_i - \E(d_i)|= O_p( (n\log n)^{1/2} ).
\]
It verifies Condition \ref{condition-diff-a}, where $h_{n1}=1$.
Similarly, by applying \citeauthor{Hoeffding:1963}'s inequality to the sum $\sum_{j<i} a_{ij}z_{ijk}$,
we have
\[
\| Q(\beta^*, \gamma^*) \| = O_p ( \kappa_n  n(\log n)^{1/2}).
\]
So $h_{n2}=\kappa_n(n\log n)^{-1/2}$ in Condition \ref{condition-Q}.
Let $\lambda_{n}$ be the smallest eigenvalue of $\bar{H}(\beta^*, \gamma^*)$. Then Condition \ref{condition-H} holds with $h_{n3}=\lambda_n$.
So by Theorem \ref{Theorem:con}, we have the following corollary.

\begin{corollary}
If
\[
\lambda_n^2\kappa_n^6 e^{24\|\beta^*\|_\infty + 12\kappa_n \|\gamma^*\|_\infty} \sqrt{\frac{\log n}{n}} = o(1),
\]
then $\|\widehat{\gamma}-\gamma^*\|_\infty = o_p(1)$ and $\|\widehat{\beta} - \beta^*\|_\infty = o_p(1)$.
\end{corollary}

Since $a_{ij}$'s ($j<i$) are independent, it is easy to show the central limit theorem for $d_i$ and $N^{-1/2}\sum_{j<i} \tilde{s}_{ij}(\beta, \gamma)$ as given in \cite{su-qian2018}
and \cite{Graham2017} respectively. So by Theorems \ref{theorem-central-b} and \ref{Theorem-central-a}, the central limit theorem holds for $\widehat{\gamma}$ and $\widehat{\beta}$.
See  \cite{su-qian2018} and \cite{Graham2017} for details.

\subsection{The Poisson model}
We consider the nonnegative integer weight for each edge, i.e., $a_{ij}\in \{0, 1, \ldots\}$.
We assume that all edges are independently distributed as Poisson random variables.
The Poisson model is
\[
\P(a_{ij}=k) = \frac{ \lambda_{ij}^k }{k!} e^{-\lambda_{ij}},
\]
where $\lambda_{ij}=  e^{z_{ij}^\top \gamma + \beta_i + \beta_j }$.
We will carry out simulations under this model in next section.
The moment equations are
\begin{equation}\label{eq:likelihood-binary}
\begin{array}{c}
d_i  =  \sum_{j\neq i}  e^{z_{ij}^\top \gamma + \beta_i + \beta_j } ,~~~i=1,\ldots, n, \\
\sum_{j<i} z_{ij}a_{ij} = \sum_{j<i}   z_{ij}e^{z_{ij}^\top \gamma + \beta_i + \beta_j },
\end{array}
\end{equation}
which are identical to the maximum likelihood equations.
Here, the $\mu$ function is $\mu(x)=e^x$.
Define
\[
q_n := \sup_{i,j} | \beta_i + \beta_j + z_{ij}^\top \gamma |.
\]
So $b_{ni}$'s ($i=0, \ldots, 3$) in inequalities \eqref{ineq-mu-keya}, \eqref{ineq-mu-keyb} and \eqref{ineq-mu-keyc} are
\[
b_{n0} = e^{-q_n}, ~~ b_{n1}= e^{q_n}, ~~ b_{n2} = e^{q_n}, ~~ b_{n3} = e^{q_n}.
\]
Lemma 8 in the supplement material shows that $h_{n1}=e^{2q_n}$ in Condition \ref{condition-diff-a}.
Similar to the lines of arguments for proving Lemma 8, we have $h_{n2}= e^{2q_n}/n^{1/2}$.
Let $\lambda_{n}$ be the smallest eigenvalue of $\bar{H}(\beta^*, \gamma^*)$. Then Condition \ref{condition-H} holds with $h_{n3}=\lambda_n$.
So by Theorem \ref{Theorem:con}, we have the following corollary.

\begin{corollary}
If
\[
\lambda_n^2\kappa_n^6 e^{28q_n} \sqrt{\frac{\log n}{n}} = o(1),
\]
then $\|\widehat{\gamma}-\gamma^*\|_\infty = o_p(1)$ and $\|\widehat{\beta} - \beta^*\|_\infty = o_p(1)$.
\end{corollary}

Note that $d_i=\sum_{j\neq i}a_{ij}$ is a sum of $n-1$ independent Poisson random variables.
Since $v_{ij} = \E a_{ij} = \lambda_{ij}$, we have
\[
e^{-q_n} \le v_{ij}= e^{ \beta_i + \beta_j + z_{ij}^\top \gamma }
\le e^{q_n},~~1\le i< j\le n.
 \]
By using the Stein-Chen identity [\cite{Stein1972, chen1975}] for the Poisson distribution, it is easy to verify that
\begin{equation}\label{eqn:poisson:exp}
\E (a_{ij}^3) = \lambda_{ij}^3 + 3\lambda_{ij}^2 + \lambda_{ij}.
\end{equation}
It follows
\[
\frac{\sum_{j\neq i} \E (a_{ij}^3) }{ v_{ii}^{3/2} } \le \frac{ (n-1)e^{q_n} }{ (n-1)^{3/2} e^{-q_n} }
= O( \frac{ e^{4q_n} }{ n^{1/2} }).
\]
If $e^{4q_n}  = o( n^{1/2} )$, then the above expression goes to zero.
This shows that the condition for the Lyapunov's central limit theorem holds.
Therefore, $v_{ii}^{-1/2} \{d_i - \E(d_i)\}$ is asymptotically standard normal under the condition $e^{4q_n}  = o( n^{1/2} )$.
When considering the asymptotic behaviors of the vector $(d_1, \ldots, d_r)$ with a fixed $r$, one could replace the degrees $d_1, \ldots, d_r$ by the independent random variables
$\tilde{d}_i=d_{i, r+1} + \ldots + d_{in}$, $i=1,\ldots,r$.
Therefore, we have the following lemma.

\begin{lemma}\label{lemma:central:poisson}
If $e^{4q_n}  = o( n^{1/2} )$, then as $n\to\infty$: \\
(1)For any fixed $r\ge 1$,  the components of $(d_1 - \E (d_1), \ldots, d_r - \E (d_r))$ are
asymptotically independent and normally distributed with variances $v_{11}, \ldots, v_{rr}$,
respectively. \\
(2)More generally, $\sum_{i=1}^n c_i(d_i-\E(d_i))/\sqrt{v_{ii}}$ is asymptotically normally distributed with mean zero
and variance $\sum_{i=1}^\infty c_i^2$ whenever $c_1, c_2, \ldots$ are fixed constants and the latter sum is finite.
\end{lemma}

Part (2) follows from part (1) and the fact that
\[
\lim_{r\to\infty} \limsup_{t\to\infty}Var( \sum_{k=r+1}^n c_i \frac{ d_i - \E (d_i) }{\sqrt{v_{ii}}})=0
\]
by Theorem 4.2 of \cite{Billingsley:1995}. To prove the above equation, it suffices to show that the eigenvalues of
the covariance matrix of $(d_i - \E (d_i))/v_{ii}^{1/2}$, $i=r+1, \ldots, n$ are bounded by 2 (for all $r<n$).
This is true by the well-known Perron-Frobenius theory:
if $A$ is a symmetric positive definite matrix with diagonal elements equaling to $1$, and with negative off-diagonal elements,
then its largest eigenvalue is less than $2$.

By \eqref{eqn:poisson:exp}, we have $\E a_{ij} \le 3\lambda_{ij}^3$.
Note that $\tilde{s}_{\gamma_{ij}} (\beta, \gamma)$ can be rewritten as
\[
\tilde{s}_{\gamma_{ij}} (\beta, \gamma)= \tilde{a}_{ij} ( z_{ij} -  V_{\gamma\beta}(\beta, \gamma)[V(\beta, \gamma)]^{-1}T_{ij} ).
\]
Let $\tilde{z}_{ij} = z_{ij} -  V_{\gamma\beta}(\beta^*, \gamma^*)[V(\beta^*, \gamma^*)]^{-1}T_{ij}$.
Once again, by applying Lyapunov's central limit theorem, we have the following lemma.

\begin{lemma}\label{lemma:th4-b}
For any nonzero vector $c=(c_1, \ldots c_p)^\top$, if
\begin{equation}\label{eqn:lemma5:a}
\frac{ \sum_{j<i } (c^\top \tilde{z}_{ij} )^3 \lambda_{ij}^3 }{ [\sum_{j<i } (c^\top \tilde{z}_{ij} )^2 \lambda_{ij} ]^{3/2} } = o(1),
\end{equation}
then $\tilde{\Sigma}^{-1/2}  \sum_{j< i}
\tilde{s}_{\gamma_{ij}} (\beta^*, \gamma^*)$ converges in distribution to the $p$-dimensional standard normal distribution, where $\tilde{\Sigma}=\sum_{j<i} \lambda_{ij} \tilde{z}_{ij} \tilde{z}_{ij}^\top$.
\end{lemma}

In view of Lemmas \ref{lemma:central:poisson} and \ref{lemma:th4-b},  the following corollary is a consequence of Theorems \ref{theorem-central-b} and \ref{Theorem-central-a}.

\begin{corollary}\label{corollary:poisson:central}
If \eqref{eqn:lemma5:a} holds and
\[
\lambda_n^2\kappa_n^6 e^{28q_n} \frac{(\log n)^{3/2}}{n^{1/2}}  = o(1),
\]
then:
(1) $N^{1/2} \overline{\Sigma}^{-1/2}(\hat{\gamma}-\gamma^*)$ converges in distribution to multivariate normal distribution with mean $\overline{\Sigma}^{-1/2}\bar{H}^{-1}B_*$  and covariance $I_p$,
where $I_p$ is the identity matrix, where $\bar{\Sigma}= N^{-1}\bar{H}^{-1} \tilde{\Sigma} \bar{H}^{-1}$; \\
(2) for a fixed $r$,  the vector  $(v_{11}^{1/2}( \hat{\beta}_1 - \beta_1^*), \ldots, v_{rr}^{1/2}( \hat{\beta}_r - \beta_r^*)$ converges in distribution to the $r$-dimensional standard normal distribution.
\end{corollary}

\section{Numerical Studies}
\label{section:simulation}
In this section, we evaluate the asymptotic results of the moment estimator under the Poisson model through simulation studies and a real data example.
The simulations for the binary weight with the logistic distribution were carried out in \cite{Graham2017}. We don't repeat here.

\subsection{Simulation studies}

We set the parameter values to be a linear form, i.e.,
$\alpha_{i}^* = (i-1)L/(n-1)$ for $i=1, \ldots, n$.
We considered four different values for $L$ as $L\in \{0 , \log(\log n), (\log n)^{1/2}, \log n \}$.
By allowing $\alpha^*$ to grow with $n$, we intended to assess the asymptotic properties under different asymptotic regimes.
Each node had two covariates $X_{i1}$ and $X_{i2}$. Specifically,
$X_{i1}$ took values positive one or negative one with equal probability and $X_{i2}$ came from a $Beta(2,2)$ distribution.
All covariates were independently generated.
The edge-level covariate $z_{ij}$ between nodes $i$ and $j$ took the form: $z_{ij}=(x_{i1}*x_{j1}, |x_{i2}-x_{j2}|)^\top$.
For the homophily parameter, we set  $\gamma^*=(0.5, 1)^\top$.
Thus, the homophily effect of the network is determined by a weighted sum of the similarity measures of the two covariates between two nodes.

By Corollary \ref{corollary:poisson:central}, given any pair $(i,j)$, $\hat{\xi}_{i,j} = [\hat{\beta}_i-\hat{\beta}_j-(\beta_i^*-\beta_j^*)]/(1/\hat{v}_{i,i}+1/\hat{v}_{j,j})^{1/2}$
converges in distribution to the standard normality, where $\hat{v}_{i,i}$ is the estimate of $v_{i,i}$
by replacing $(\beta^*, \gamma^*)$ with $(\widehat{\beta}, \widehat{\gamma})$.
Therefore, we assessed the asymptotic normality of $\hat{\xi}_{i,j}$ using the quantile-quantile (QQ) plot.
Further, we also recorded the coverage probability of the 95\% confidence interval, the length of the confidence interval.
The coverage probability and the length of the confidence interval of $\widehat{\gamma}$ were also reported. Finally,
each simulation was repeated $10,000$ times.

\begin{figure}[!htb]
\centering
\includegraphics[ width=6.5in, angle=0]{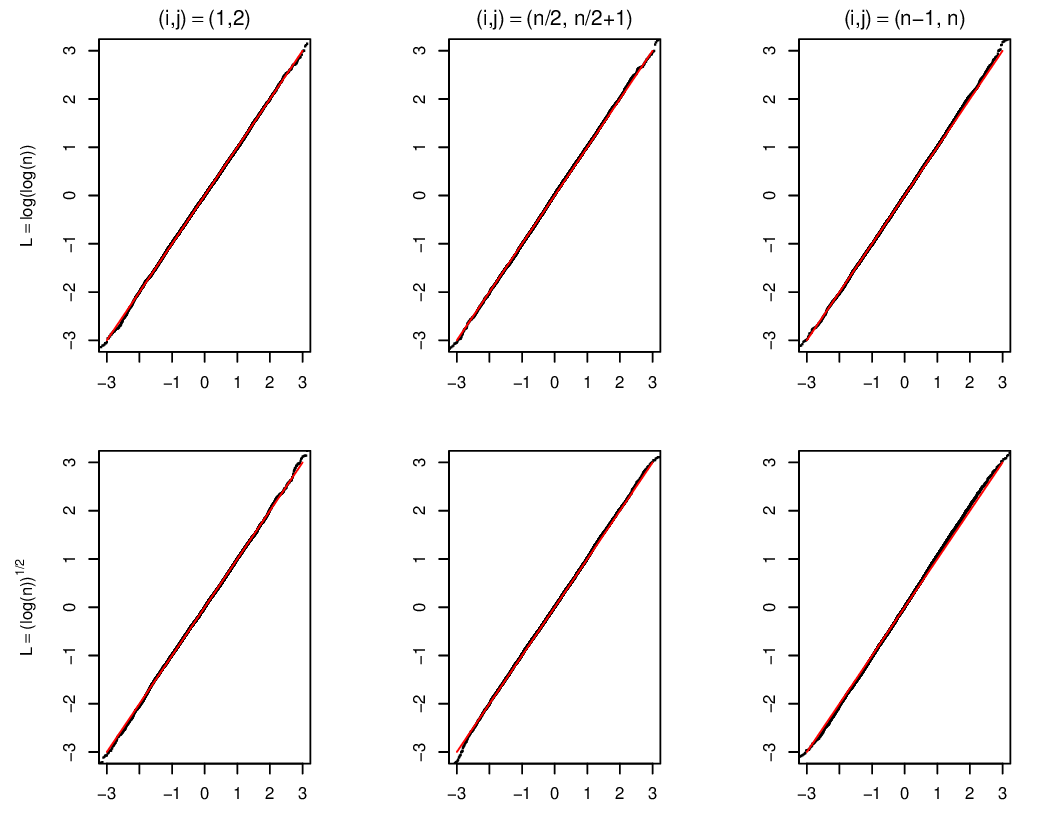}
\caption{The QQ plots of $\hat{\xi}_{i,j}$ (n=100). }
\label{figure-qq}
\end{figure}

We did simulations with network sizes $n=100$ and $n=200$ and found that the QQ-plots for these two network sizes were similar.
Therefore, we only show the QQ-plots for $n=100$ to save space.
Further, the QQ-plots for $L=0$ and $L=\log(\log n)$  are similar.
Also, for $L=(\log n)^{1/2}$ and $L=\log n$, they are similar.
Therefore we only show those for $L=\log (\log n)$ and $L=\log n$  in Figure \ref{figure-qq}.
In this figure, the horizontal and vertical axes are the theoretical and empirical quantiles, respectively,
and the straight lines correspond to the reference line $y=x$.
In Figure \ref{figure-qq}, when $L=\log (\log n)$, the empirical quantiles coincide well with the theoretical ones.
When $L=(\log n)^{1/2}$, the empirical quantiles have a little derivation from the theoretical ones in the upper tail of the right bottom subgraph.
These figures show that there may be large space for improvement on the growing rate of $\| \beta\|_\infty$ in the conditions in Corollary \ref{corollary:poisson:central}.

Table \ref{Table:alpha} reports the coverage probability of the 95\% confidence interval for $\beta_i - \beta_j$ and the length of the confidence interval.
As we can see, the length of the confidence interval decreases as $n$ increases, which qualitatively agrees with the theory.
The coverage frequencies are all close to the nominal level $95\%$. On the other hand, the length of the confidence interval decreases as $L$ increases.
It seems  a little unreasonable. Actually,  the theoretical length of the $95\%$ confidence interval is $(1/v_{ii} + v_{jj})^{1/2}$ multiple by
a constant factor. Since $v_{ii}$ is a sum of a set of exponential items,  it becomes quickly larger as $L$ increases. As a result,
the length of confidence interval decreases as long as the estimates are close to the true values.  The simulated coverage probability results shows that
the estimates are very good.
So, this phenomenon that the length of confidence interval decreases in Table \ref{Table:alpha}, also agrees with the theory.

{\renewcommand{\arraystretch}{1}
\begin{table}[!h]\centering
\caption{The reported values are the coverage frequency ($\times 100\%$) for $\beta_i-\beta_j$ for a pair $(i,j)$ / the length of the confidence interval($\times 10$).}
\label{Table:alpha}
\begin{tabular}{ccccccc}
\hline
n       &  $(i,j)$ & $L=0$ & $L=\log(\log n)$ & $L=(\log n)^{1/2}$ & $L=\log n$ \\
\hline
100         &$(1,2)   $&$94.56 / 4.60 $&$ 95.08 / 2.97 $&$ 94.80 / 2.42 $&$ 94.69 / 0.97  $ \\
            &$(50,51) $&$94.72 / 4.60 $&$ 94.93 / 2.04 $&$ 94.89 / 1.43 $&$ 94.83 / 0.31 $ \\
            &$(99,100)$&$95.12 / 4.60 $&$ 94.41 / 1.40 $&$ 94.38 / 0.85 $&$ 94.13 / 0.10$ \\
&&&&&&\\
200         &$(1,2)     $&$95.20 / 3.24 $&$ 94.79 / 2.01 $&$ 94.76 / 1.63 $&$ 95.09 / 0.52 $ \\
            &$(100,101) $&$95.03 / 3.24 $&$ 94.75 / 1.33 $&$ 94.91 / 0.92 $&$ 95.47 / 0.14 $ \\
            &$(199,200) $&$94.58 / 3.24 $&$ 95.05 / 0.88 $&$ 94.63 / 0.52 $&$ 93.90 / 0.04$ \\
\hline
\end{tabular}
\end{table}
}

Table \ref{Table:gamma} reports the coverage frequencies for the estimate $\widehat{\gamma}$
and bias corrected estimate  $\widehat{\gamma}_{bc}$ 
at the nominal level $95\%$,
and the standard error.
As we can see, the differences between the coverage frequencies with  uncorrected estimates and bias corrected estimates are very small, less than $0.1$.
All coverage frequencies are very close to the nominal level. This implies that the bias is very small in our simulation design.

{\renewcommand{\arraystretch}{1}
\begin{table}[!htbp]\centering
\caption{
The reported values are the coverage frequency ($\times 100\%$) for $\gamma_i$ for $i$ / length ($\times 10$) of confidence interval ($\bs{\gamma}^*=(0.5, 1)^\top$).
}
\label{Table:gamma}
\begin{tabular}{cclllcc}
\hline
$n$     &   $\bs{\widehat{\gamma}}$  & $L=0$ & $L=\log(\log n)$ & $L=(\log n)^{1/2}$ & $L=\log n$ \\
\hline
$100$   & $\hat{\gamma}_1$            &$95.13 / 0.52 $&$ 95.25 / 0.22 $&$ 94.92 / 0.15 $&$ 95.04 / 0.02 $ \\

        & $\hat{\gamma}_{bc, 1}$      &$95.11 / 0.52 $&$ 95.25 / 0.22 $&$ 94.92 / 0.15 $&$ 95.04 / 0.02 $ \\

        & $\hat{\gamma}_2$            &$94.98 / 3.08 $&$ 95.28 / 1.31 $&$ 95.00 / 0.88 $&$ 95.06 / 0.15  $ \\

        & $\hat{\gamma}_{bc, 2}$      &$94.93 / 3.08 $&$ 95.29 / 1.31 $&$ 95.02 / 0.88 $&$ 95.06 / 0.15  $\\

$200$   & $\hat{\gamma}_1$            &$94.87 / 0.26 $&$ 95.49 / 0.10 $&$ 95.07 / 0.07 $&$ 94.92 / 0.007 $ \\
        & $\hat{\gamma}_{bc, 1}$      &$94.87 / 0.26 $&$ 95.47 / 0.10 $&$ 95.08 / 0.07 $&$ 94.91 / 0.007 $\\

        & $\hat{\gamma}_2$            &$95.31 / 1.52 $&$ 95.12 / 0.59 $&$ 94.97 / 0.39 $&$ 94.49 / 0.041 $ \\
        & $\hat{\gamma}_{bc, 2}$      &$95.31 / 1.52 $&$ 95.12 / 0.59 $&$ 94.95 / 0.39 $&$ 94.49 / 0.041 $\\

\hline
\end{tabular}
\end{table}
}

\subsection{A real data example}


We use the Enron email dataset as an example analysis [\cite{Cohen2004}], available from \url{https://www.cs.cmu.edu/~enron/}.  
This dataset was released by William Cohen at Carnegie Mellon University and is  now the May 7, 2015 Version of dataset,
which is widely accepted by many researchers.
The Enron email dataset is valuable because it is one of the very few collections of organizational emails that are publicly available.
The reason that other datasets are not public, is because of privacy concerns.
The Enron email data was acquired and made public by the Federal Energy Regulatory Commission during its investigation into fraudulent accounting practices.
Some of the emails have been deleted upon requests from affected employees.
However, the raw data is messy and needs to be cleaned before any analysis is conducted.
\cite{Zhou2007} applied data cleaning strategies to compile the Enron email dataset.
We use \citeauthor{Zhou2007}'s cleaned data for the subsequent analysis.
The resulting data comprises $21,635$ messages sent between $156$ employees
with their covarites information. There are $6,650$ messages having more than one recipient across their `To', `CC' and
`BCC' fields, with a few messages having more than 50 recipients. For our analysis,
we exclude messages with more than ten recipients, which is a subjectively chosen cut-off that avoids
emails sent en masse to large groups. Each employee has three categorical variables: departments of these employees (Trading, Legal, Other),
the genders (Male, Female) and seniorities (Senior, Junior). Employees are labelled from $1$ to $156$.
The $3$-dimensional covariate vector $z_{ij}$ of edge $(i,j)$ is formed by using a homophilic matching function between these $3$ covariates of two employees $i$ and $j$, i.e.,
if $x_{ik}$ and $x_{jk}$ are equal, then $z_{ijk}=1$; otherwise $z_{ijk}=0$.


{\renewcommand{\arraystretch}{1}
\begin{table}[!hbt]\centering
\scriptsize
\caption{The estimates of $\beta_i$ and their standard errors in the Enron email dataset.}
\label{Table:alpha:real}
\begin{tabular}{cccc c cccc c cccc c cccc}
\hline
Node & $d_i$  &  $\hat{\beta}_i$ & $\hat{\sigma}_i$ & Node & $d_i$  &  $\hat{\beta}_i$ & $\hat{\sigma}_i$ &  Node & $d_i$  &  $\hat{\beta}_i$ & $\hat{\sigma}_i$
& Node & $d_i$  &  $\hat{\beta}_i$ & $\hat{\sigma}_i$ \\
\hline
1 &$ 723 $&$ 1.03 $&$ 0.37 $& 41 &$ 309 $&$ 0.15 $&$ 0.57 $& 79 &$ 309 $&$ -0.46 $&$ 0.79 $& 117 &$ 1176 $&$ 1.49 $&$ 0.29 $ \\
2 &$ 67 $&$ -1.36 $&$ 1.22 $& 42 &$ 281 $&$ 0.08 $&$ 0.6 $& 80 &$ 281 $&$ -0.08 $&$ 0.65 $& 118 &$ 398 $&$ 0.4 $&$ 0.5 $ \\
3 &$ 275 $&$ 0.03 $&$ 0.6 $& 43 &$ 690 $&$ 0.96 $&$ 0.38 $& 81 &$ 690 $&$ 0.32 $&$ 0.53 $& 119 &$ 369 $&$ 0.35 $&$ 0.52 $ \\
4 &$ 1202 $&$ 1.54 $&$ 0.29 $& 44 &$ 234 $&$ -0.13 $&$ 0.65 $& 82 &$ 234 $&$ 0.32 $&$ 0.52 $& 120 &$ 2673 $&$ 2.33 $&$ 0.19 $ \\
5 &$ 678 $&$ 0.94 $&$ 0.38 $& 45 &$ 704 $&$ 1 $&$ 0.38 $& 83 &$ 704 $&$ -1.45 $&$ 1.27 $& 121 &$ 571 $&$ 0.75 $&$ 0.42 $ \\
6 &$ 249 $&$ -0.07 $&$ 0.63 $& 46 &$ 952 $&$ 1.27 $&$ 0.32 $& 84 &$ 952 $&$ -0.74 $&$ 0.89 $& 122 &$ 2174 $&$ 2.15 $&$ 0.21 $ \\
7 &$ 375 $&$ 0.35 $&$ 0.52 $& 47 &$ 998 $&$ 1.38 $&$ 0.32 $& 85 &$ 998 $&$ 0.72 $&$ 0.43 $& 123 &$ 343 $&$ 0.26 $&$ 0.54 $ \\
8 &$ 40 $&$ -1.88 $&$ 1.58 $& 48 &$ 686 $&$ 0.99 $&$ 0.38 $& 86 &$ 686 $&$ -2.04 $&$ 1.71 $& 124 &$ 115 $&$ -0.8 $&$ 0.93 $ \\
9 &$ 428 $&$ 0.48 $&$ 0.48 $& 49 &$ 1224 $&$ 1.54 $&$ 0.29 $& 87 &$ 1224 $&$ -0.31 $&$ 0.71 $& 125 &$ 195 $&$ -0.29 $&$ 0.72 $ \\
10 &$ 95 $&$ -1.01 $&$ 1.03 $& 50 &$ 141 $&$ -0.63 $&$ 0.84 $& 88 &$ 141 $&$ -1.29 $&$ 1.16 $& 126 &$ 102 $&$ -0.96 $&$ 0.99 $ \\
11 &$ 231 $&$ -0.12 $&$ 0.66 $& 51 &$ 101 $&$ -0.95 $&$ 1 $& 89 &$ 101 $&$ -1.31 $&$ 1.17 $& 127 &$ 180 $&$ -0.4 $&$ 0.75 $ \\
12 &$ 31 $&$ -2.16 $&$ 1.8 $& 52 &$ 1 $&$ -5.57 $&$ 10 $& 90 &$ 1 $&$ 0.52 $&$ 0.48 $& 128 &$ 67 $&$ -1.39 $&$ 1.22 $ \\
13 &$ 85 $&$ -1.15 $&$ 1.08 $& 53 &$ 1138 $&$ 1.46 $&$ 0.3 $& 91 &$ 1138 $&$ 1.17 $&$ 0.35 $& 129 &$ 185 $&$ -0.38 $&$ 0.74 $ \\
14 &$ 53 $&$ -1.62 $&$ 1.37 $& 54 &$ 66 $&$ -1.41 $&$ 1.23 $& 92 &$ 66 $&$ 1.59 $&$ 0.28 $& 130 &$ 1798 $&$ 1.96 $&$ 0.24 $ \\
15 &$ 182 $&$ -0.36 $&$ 0.74 $& 55 &$ 155 $&$ -0.5 $&$ 0.8 $& 93 &$ 155 $&$ -1.02 $&$ 1.03 $& 131 &$ 3157 $&$ 2.5 $&$ 0.18 $ \\
16 &$ 26 $&$ -2.34 $&$ 1.96 $& 56 &$ 266 $&$ 0.02 $&$ 0.61 $& 94 &$ 266 $&$ -1.49 $&$ 1.3 $& 132 &$ 98 $&$ -0.96 $&$ 1.01 $ \\
17 &$ 702 $&$ 0.98 $&$ 0.38 $& 57 &$ 555 $&$ 0.76 $&$ 0.42 $& 95 &$ 555 $&$ 0.94 $&$ 0.38 $& 133 &$ 57 $&$ -1.5 $&$ 1.32 $ \\
18 &$ 182 $&$ -0.36 $&$ 0.74 $& 58 &$ 423 $&$ 0.47 $&$ 0.49 $& 96 &$ 423 $&$ -2.22 $&$ 1.86 $& 134 &$ 106 $&$ -0.93 $&$ 0.97 $ \\
19 &$ 122 $&$ -0.78 $&$ 0.91 $& 59 &$ 3715 $&$ 2.69 $&$ 0.16 $& 97 &$ 3715 $&$ -1.88 $&$ 1.58 $& 135 &$ 182 $&$ -0.39 $&$ 0.74 $ \\
20 &$ 4637 $&$ 2.97 $&$ 0.15 $& 60 &$ 298 $&$ 0.14 $&$ 0.58 $& 98 &$ 298 $&$ 0.79 $&$ 0.41 $& 136 &$ 79 $&$ -1.19 $&$ 1.13 $ \\
21 &$ 14 $&$ -2.96 $&$ 2.67 $& 61 &$ 1832 $&$ 1.97 $&$ 0.23 $& 99 &$ 1832 $&$ -1.96 $&$ 1.62 $& 137 &$ 676 $&$ 0.96 $&$ 0.38 $ \\
22 &$ 44 $&$ -1.8 $&$ 1.51 $& 62 &$ 65 $&$ -1.41 $&$ 1.24 $& 100 &$ 65 $&$ 0.31 $&$ 0.53 $& 138 &$ 2340 $&$ 2.23 $&$ 0.21 $ \\
23 &$ 135 $&$ -0.69 $&$ 0.86 $& 63 &$ 419 $&$ 0.46 $&$ 0.49 $& 101 &$ 419 $&$ -0.19 $&$ 0.67 $& 139 &$ 3 $&$ -4.5 $&$ 5.77 $ \\
24 &$ 826 $&$ 1.15 $&$ 0.35 $& 64 &$ 68 $&$ -1.37 $&$ 1.21 $& 102 &$ 68 $&$ -0.34 $&$ 0.72 $& 140 &$ 208 $&$ -0.2 $&$ 0.69 $ \\
25 &$ 135 $&$ -0.64 $&$ 0.86 $& 65 &$ 1159 $&$ 1.48 $&$ 0.29 $& 103 &$ 1159 $&$ -1.48 $&$ 1.3 $& 141 &$ 56 $&$ -1.56 $&$ 1.34 $ \\
26 &$ 668 $&$ 0.95 $&$ 0.39 $& 66 &$ 170 $&$ -0.45 $&$ 0.77 $& 104 &$ 170 $&$ -1.04 $&$ 1.03 $& 142 &$ 241 $&$ -0.08 $&$ 0.64 $ \\
27 &$ 644 $&$ 0.88 $&$ 0.39 $& 67 &$ 815 $&$ 1.13 $&$ 0.35 $& 105 &$ 815 $&$ -1.65 $&$ 1.39 $& 143 &$ 645 $&$ 0.88 $&$ 0.39 $ \\
28 &$ 20 $&$ -2.59 $&$ 2.24 $& 68 &$ 112 $&$ -0.87 $&$ 0.94 $& 106 &$ 112 $&$ -1.3 $&$ 1.19 $& 144 &$ 540 $&$ 0.71 $&$ 0.43 $ \\
29 &$ 190 $&$ -0.34 $&$ 0.73 $& 69 &$ 707 $&$ 0.99 $&$ 0.38 $& 107 &$ 707 $&$ -1.38 $&$ 1.21 $& 145 &$ 1080 $&$ 1.43 $&$ 0.3 $ \\
30 &$ 99 $&$ -0.97 $&$ 1.01 $& 70 &$ 33 $&$ -2.09 $&$ 1.74 $& 108 &$ 33 $&$ -1.32 $&$ 1.18 $& 146 &$ 67 $&$ -1.39 $&$ 1.22 $ \\
31 &$ 60 $&$ -1.47 $&$ 1.29 $& 71 &$ 136 $&$ -0.68 $&$ 0.86 $& 109 &$ 136 $&$ 1.12 $&$ 0.35 $& 147 &$ 440 $&$ 0.51 $&$ 0.48 $ \\
33 &$ 241 $&$ -0.11 $&$ 0.64 $& 72 &$ 788 $&$ 1.12 $&$ 0.36 $& 110 &$ 788 $&$ -0.95 $&$ 0.99 $& 148 &$ 165 $&$ -0.49 $&$ 0.78 $ \\
34 &$ 996 $&$ 1.35 $&$ 0.32 $& 73 &$ 179 $&$ -0.41 $&$ 0.75 $& 111 &$ 179 $&$ -1.07 $&$ 1.07 $& 149 &$ 588 $&$ 0.8 $&$ 0.41 $ \\
35 &$ 96 $&$ -0.98 $&$ 1.02 $& 74 &$ 720 $&$ 1 $&$ 0.37 $& 112 &$ 720 $&$ -0.03 $&$ 0.62 $& 150 &$ 38 $&$ -1.95 $&$ 1.62 $ \\
36 &$ 97 $&$ -1.02 $&$ 1.02 $& 75 &$ 313 $&$ 0.15 $&$ 0.57 $& 113 &$ 313 $&$ 1.21 $&$ 0.33 $& 151 &$ 1330 $&$ 1.65 $&$ 0.27 $ \\
38 &$ 564 $&$ 0.74 $&$ 0.42 $& 76 &$ 184 $&$ -0.38 $&$ 0.74 $& 114 &$ 184 $&$ -0.04 $&$ 0.62 $& 152 &$ 120 $&$ -0.81 $&$ 0.91 $ \\
39 &$ 711 $&$ 0.98 $&$ 0.38 $& 77 &$ 358 $&$ 0.32 $&$ 0.53 $& 115 &$ 358 $&$ -0.06 $&$ 0.65 $& 153 &$ 219 $&$ -0.21 $&$ 0.68 $ \\
40 &$ 202 $&$ -0.29 $&$ 0.7 $& 78 &$ 137 $&$ -0.64 $&$ 0.85 $& 116 &$ 137 $&$ -0.94 $&$ 0.99 $& 154 &$ 298 $&$ 0.1 $&$ 0.58 $ \\
155 &$ 82 $&$ -1.17 $&$ 1.1$& 156 &$ 480 $&$ 0.6 $&$ 0.46$ \\
\hline
\end{tabular}

\end{table}

For our analysis, we removed the employees  ``32" and ``37" with zero degrees in this case the estimators of the corresponding degree parameters do not exist.
This leaves a connected network with $154$ nodes.
The minimum, $1/4$ quantile, median, $3/4$ quantile and maximum values of $d$ are $1$, $95$, $220$, $631$ and $4637$, respectively.
It exhibits a strong degree heterogeneity. The estimators of $\alpha_i$ with their estimated standard errors are given in Table \ref{Table:alpha:real}.
The estimates of degree parameters vary widely:
from the minimum $-4.36$ to maximum $2.97$.
We then test three null hypotheses $\beta_2=\beta_3$, $\beta_{76}=\beta_{77}$ and $\beta_{151}=\beta_{154}$, using
the homogeneity test statistics $\hat{\xi}_{i,j} = |\hat{\beta}_i-\hat{\beta}_j|/(1/\hat{v}_{i,i}+1/\hat{v}_{j,j})^{1/2}$.
The obtained $p$-values turn out to be $1.7\times 10^{-24}$, $1.8\times 10^{-4}$ and $6.2\times 10^{-23}$, respectively,
confirming the need to assign one parameter to each node to characterize the heterogeneity of degrees.

The estimated covariate effects, their  bias corrected estimates, their standard errors, and their $p$-values under the null of having no effects are reported in Table \ref{Table:gamma:realdata}.
From this table, we can see that the estimates and bias corrected estimates are the same, indicating that
the bias effect is very small in the Poisson model and it corroborates the findings of simulations.
The variables department and seniority are significant while gender is not significant.
This indicates that the gender has no significant influence on the formation of organizational emails.
The coefficient of variable department is positive,
implying that a common value increases the probability of two employees in the same department to have more email connections.
On the other hand, the coefficient of variable seniority is negative, indicating that
two employees in the same seniority have less emails than those with unequal seniorities. This makes sense intuitively.

{\renewcommand{\arraystretch}{1}
\begin{table}[h]\centering
\caption{The estimators of $\gamma_i$, the corresponding bias corrected estimators, the standard errors, and the $p$-values under the null $\gamma_i=0$ ($i=1, 2, 3$) for Enron email data.}
\label{Table:gamma:realdata}
\begin{tabular}{ccc ccc cc c}
\hline
Covariate 
&  $\hat{\gamma}_i$ & $\hat{\gamma}_{bc, i}$ & $\hat{\sigma}_i$ &$p$-value  \\
\hline
Department          
&  $ 0.167 $&$ 0.167 $&$ 1.13$ &$ <0.001$\\
Gender          
&  $-0.006 $&$ -0.006 $&$ 1.27$&$ 0.62$\\
Seniority       
&  $-0.203 $&$ -0.203 $&$ 1.09$&$ <0.001$\\
\hline
\end{tabular}
\end{table}
{\renewcommand{\arraystretch}{1}

\section{Summary and discussion}
\label{section:sd}
We have present the moment estimation for inferring the degree parameters and homophily parameter in model \eqref{model}
that only specifies the marginal distribution. We establish
the consistency of the moment estimator under several conditions and also derive its asymptotic representation.
It is worth noting that the conditions imposed on $b_{n0}$ and $b_{n1}$ may not be best possible.  In particular, the conditions in Theorems \ref{theorem-central-b} and \ref{Theorem-central-a} seem stronger than those needed for the consistency. Note that the asymptotic behavior of the MLE depends not only on $b_{n0}$ and $b_{n1}$, but also on the configuration of the parameters.
We will investigate this in the future.

Throughout the paper, we assume that $\max_{i,j} \| z_{ij} \|_\infty \le \kappa_n$.
Conditions imposed in the theorems imply that $\kappa_n$ can be allowed to increase only with a slow rate.
What can be said when some of $\|z_{ij}\|_\infty$'s are large? For
example, some of the covariates information for edges may increase with a fast rate.
If the proportion of large values of $\|z_{ij}\|_\infty$'s is bounded, then this will have little effect
on the moment estimators when $n$ is large, so that the consistency
and asymptotic representation still hold.
 A more interesting case is when the proportion of large $\|z_{ij}\|_\infty$'s
 is not bounded,  whether there are any asymptotic properties of the moment estimator. We plan to
investigate this and other related situations in the future.


In this paper, we make an edge independence assumption. When edges are not independent,
our main results (Theorems \ref{Theorem:con}, \ref{Theorem-central-a} and \ref{theorem-central-b}) still hold
as long as Conditions \ref{condition-diff-a}, \ref{condition-Q} and \ref{condition-H} satisfy.
In fact, the edge independence assumption are not directly used through checking our proofs.
And this assumption is only used in our applications to verify Condition \ref{condition-diff-a} and to derive the central limit theorem.
In the edge dependence case, there are also a lot of Hoeffding-type exponential tail inequalities [e.g., \cite{Delyon:2009,Roussas1996,Ioannides1999423}]
and cental limit theorems for sums of a sequence of random variables (e.g. \cite{cocke1972,cox1984}) to apply.
We hope that the results developed here can be applied to more general network models.
Finally, we mention some results for network models with dependence edges.
If the exponential random graph models include network configurations such as $k$-stars and triangles are included as sufficient statistics.
then such models incur the problem of model degeneracy in the sense of \cite{Handcock-2003}, in which almost all realized
graphs essentially have no edges or are complete, completely skipping all intermediate structures.
\cite{Chatterjee:Diaconis:2013} shew that most realizations from many ERGMs
 look like the results of a simple Erdos-Renyi model and
gave a rigorous proof of the degeneracy observed in the edge-triangle model. 
\cite{Yin:2015} further gave an explicit characterization of the degenerate tendency as a function of the parameters.
On the other hand, the MLE in ERGMs with dependent structures also incur problematic properties.
\cite{Shalizi:Rinaldo:2013} demonstrated that the MLE is not consistent.
On the other hand, some refined network statistics such as ``alternating $k$-stars", ``alternating $k$-triangles" and so on
in \cite{Robins.et.al.2007b} are proposed, but the theoretical properties of the model are still unknown.

\section{Appendix}
\label{section:appendix}
\subsection{Preliminaries}

In this section, we present two results that will be used in the proofs.
The first is on the approximation error of using $S$ to approximate the inverse of $V$ belonging to the matrix class $\mathcal{L}_n(m, M)$,
where $V=(v_{ij})_{n\times n}$ and $S=\mathrm{diag}(1/v_{11}, \ldots, 1/v_{nn})$.
\cite{Yan:Zhao:Qin:2015} obtain the upper bound of the approximation error stated below,
which has an order $n^{-2}$.

\begin{proposition}[Proposition 1 in \cite{Yan:Zhao:Qin:2015}] \label{pro:inverse:appro}
If $V \in \mathcal{L}_n(m, M)$, then the following holds: 
\begin{equation}\label{O-upperbound}
\|V^{-1} - S \|_{\max}=O\left(\frac{M^2}{n^2m^3}\right).
\end{equation}
\end{proposition}

The other result is the rate of convergence for the Newton method.
There are many convergence results on the Newton method; see the book by \cite{suli:Mayers:2003} for a comprehensive survey.
We use \citeauthor{Gragg:Tapia:1974}'s \citeyearpar{Gragg:Tapia:1974} result here.

\begin{theorem}[\cite{Gragg:Tapia:1974}]\label{pro:Newton:Kantovorich}
Let $D$ be an open convex set of $\R^n$ and $F:D \to \R^n$ a differential function
with a Jacobian $F^\prime(x)$ that is Lipschitz continuous on $D$ with Lipschitz coefficient $\lambda$.
Assume that $x_0 \in D$ is such that $[ F^\prime (x_0) ]^{-1} $ exists,
\[
\| [ F^\prime (x_0 ) ]^{-1} \|_\infty  \le \aleph,~~ \| [ F^\prime (x_0) ]^{-1} F(x_0) \|_\infty \le \delta, ~~ \rho= 2 \aleph \lambda \delta \le 1,
\]
and
\[
B(x_0, t^*) \subset D, ~~ t^* = \frac{2}{\rho} ( 1 - \sqrt{1-\rho} ) \delta = \frac{ 2\delta }{ 1 + \sqrt{1+\rho} }.
\]
Then: (1) The Newton iterations $x_{k+1} = x_k - [ F^\prime (x_k) ]^{-1} F(x_k)$ exist and $x_k \in B(x_0, t^*) \subset D$ for $k \ge 0$. (2)
$x^* = \lim x_k$ exists, $x^* \in \overline{ B(x_0, t^*) } \subset D$ and $F(x^*)=0$.
\end{theorem}

\subsection{Proof of Lemma \ref{lemma-a}}
Note that $\widehat{\beta}_{\gamma}$ is the solution to
the equation $F_{\gamma}(\beta)$=0.
To prove this lemma, it is sufficient to show that the Newton-Kantovorich conditions for the function $F_{\gamma}(\beta)$ hold when
$D=B(\beta^*, \epsilon_{n1})$ and $\gamma\in B(\gamma^*, \epsilon_{n2})$, where $\epsilon_{n1}$ is a positive number and
$\epsilon_{n2}= (\log n/n)^{1/2}$. 
The following calculations are based on the event $E_n$:
\[
E_n = \{d: \max_i | d_i - \E d_i | = O( h_{n1}(n\log n)^{1/2} ) \}.
\]
In the Newton iterative step, we set the true parameter vector $\beta^*$
as the starting point $\beta^{(0)}:=\beta^*$.
By Proposition \ref{condition:matrixclass-d}, we have $F_{\gamma}(\beta^*)\in \mathcal{L}_n(b_{n0}, b_{n1})$ or $-F_{\gamma}(\beta^*)\in \mathcal{L}_n(b_{n0}, b_{n1})$
when $\beta\in B(\beta^*, \epsilon_{n1})$ and $\gamma \in B(\gamma^*, \epsilon_{n2})$.
The proofs under two cases $F_{\gamma}(\beta^*)\in \mathcal{L}_n(b_{n0}, b_{n1})$ or $-F_{\gamma}(\beta^*)\in \mathcal{L}_n(b_{n0}, b_{n1})$, are similar.
We only give the proof under the first case.

Let $V=(v_{ij})= \partial F_{\gamma}(\beta^*)/\partial \beta^\top$ and $S=\mathrm{diag}(1/v_{11}, \ldots, 1/v_{nn})$.
By Proposition \ref{pro:inverse:appro}, we have
\[
\aleph =\|V^{-1}\|_\infty \le \| V^{-1} -S\|_\infty + \| S \|_\infty = O(\frac{b_{n1}^2}{nb_{n0}^3}) + O( \frac{1}{nb_{n0}})=O(\frac{b_{n1}^2}{nb_{n0}^3}).
\]
Recall that $F_{\gamma^*}(\beta^*) = d - \E d$ and $\gamma\in B(\gamma^*, (\log n/n)^{1/2})$.
Note that the dimension $p$ of $\gamma$  is a fixed constant.
By the event $E_n$ and inequality \eqref{ineq-mu-gamma}, we have
\begin{eqnarray*}
\| F_\gamma(\beta^*) \|_\infty & \le & \| d - \E d \|_\infty + \max_i | \sum\nolimits_{j\neq i} [\mu_{ij}(\beta^*, \gamma) - \mu_{ij}(\beta^*, \gamma^*)] |  \\
& \le & (h_{n1} + p b_{n1}\kappa_n )(n\log n)^{1/2}.
\end{eqnarray*}
Repeatedly utilizing  Proposition \ref{pro:inverse:appro}, we have
\begin{eqnarray*}
\delta&=&\| [F'_\gamma(\beta^*)]^{-1}F_\gamma(\beta^*) \|_\infty \\
& \le &
n\|V^{-1} - S\| \|F_\gamma(\beta^*)\|_\infty + \max_{i}\frac{|F_{i,\gamma}(\beta^*)|}{v_{ii}} \\
& \le & \left[ O(\frac{ b_{n1}^2}{nb_{n0}^3}) + \frac{ 1}{(n-1)b_{n0}} \right] \times (h_{n1}+pb_{n1}\kappa_n)(n\log n)^{1/2}  \\
& = & O\left( \frac{ (h_{n1}+b_{n1}\kappa_n)b_{n1}^2}{b_{n0}^3}\sqrt{\frac{\log n}{n}} \right).
\end{eqnarray*}
By Proposition \ref{condition:lipschitz-c}, $F_\gamma(\beta)$ is Lipschitz continuous with Lipschitz coefficient $\lambda=4b_{n2}(n-1)$.
Therefore, if \eqref{equation:lemma-a} holds, then
\begin{eqnarray*}
\rho =2\aleph \lambda \delta & = & O(\frac{b_{n1}^2}{nb_{n0}^3})\times O( b_{n2}n ) \times O( \frac{ (h_{n1}+b_{n1}\kappa_n)b_{n1}^2}{b_{n0}^3}\sqrt{\frac{\log n}{n}} ) \\
& = & O\left( \frac{b_{n1}^4 b_{n2}(h_{n1}+b_{n1}\kappa_n)}{b_{n0}^6}\times \frac{(\log n)^{1/2}}{n^{1/2}} \right) =o(1).
\end{eqnarray*}
The above arguments verify the Newton-Kantovorich conditions.
By Theorem \ref{pro:Newton:Kantovorich}, it yields that
\[
\| \widehat{\beta}_\gamma - \beta^* \|_\infty = O\left( \frac{ b_{n1}^2(h_{n1} + b_{n1}\kappa_n)}{b_{n0}^3}\sqrt{\frac{\log n}{n}} \right).
\]
By Condition \ref{condition-diff-a}, $P(E_n)\to 1$ such that the above equation holds with probability approaching one.
It completes the proof.

\subsection{Deriving the expression of \eqref{equation:Qc-derivative}}
Note that $F(\widehat{\beta}_\gamma, \gamma)=0$. By the compound function derivation law, we have
\begin{equation}\label{equ-derivation-a}
\frac{ \partial F(\beta, \gamma) }{\partial \gamma^\top} \bigg |_{ \beta=\widehat{\beta}_\gamma, \gamma=\gamma}
 = \frac{ \partial F(\beta, \gamma) }{\partial \beta^\top}\bigg |_{ \beta=\widehat{\beta}_\gamma, \gamma=\gamma}
  \frac{\partial \widehat{\beta}_\gamma }{\gamma^\top} + \frac{\partial F(\beta, \gamma)}{\partial \gamma^\top}\bigg |_{ \beta=\widehat{\beta}_\gamma, \gamma=\gamma}=0,
\end{equation}
and
\begin{equation}\label{equ-derivation-b}
\frac{ \partial Q_c(\gamma)}{ \partial \gamma^\top} = \left.\frac{\partial Q(\beta, \gamma)}{\partial \beta^\top}\right|_{\beta=\widehat{\beta}_\gamma,\gamma=\gamma}
 \frac{\partial \widehat{\beta}_\gamma }{\gamma^\top} + \left.\frac{ \partial Q(\beta, \gamma) }{ \partial \gamma^\top}\right|_{\beta=\widehat{\beta}_\gamma,\gamma=\gamma}.
\end{equation}
Formula \eqref{equ-derivation-a} implies that
\[
\frac{ \partial\widehat{\beta}_\gamma }{ \partial \gamma^\top} = -
\left[\frac{\partial F(\beta, \gamma)}{\partial \beta^\top}\bigg |_{ \beta=\widehat{\beta}_\gamma, \gamma=\gamma} \right]^{-1}
\frac{\partial F(\beta, \gamma)}{\partial \gamma^\top}|_{ \beta=\widehat{\beta}_\gamma, \gamma=\gamma}.
\]
Substituting the above expression into \eqref{equ-derivation-b}, it yields equation \eqref{equation:Qc-derivative}.

\subsection{Proof of Theorem \ref{Theorem:con}}
We only give the proof in the case $F_\gamma^\prime(\beta)\in \mathcal{L}_n(b_{n0}, b_{n1})$. The proof in the case of that
$-F_\gamma^\prime(\beta)\in \mathcal{L}_n(b_{n0}, b_{n1})$,
is similar, and we omit it.
We construct the Newton iterative sequence to show the consistency. It is sufficient to verify the
Newton-Kantovorich conditions as in the proof of Lemma \ref{lemma-a}. 
We set $\gamma^*$ as the initial point $\gamma^{(0)}$ and $\gamma^{(k+1)}=\gamma^{(k)} - [Q_c^\prime(\gamma^{(k)})]^{-1}Q_c(\gamma^{(k)})$.

By Lemma \ref{lemma-a}, 
we have
\[
\| \widehat{\beta}_\gamma - \beta^* \|_\infty = O_p\left( \frac{ b_{n1}^2(h_{n1}+b_{n1})}{b_{n0}^3}\sqrt{\frac{\log n}{n}} \right). 
\]
This shows that $\widehat{\beta}_{\gamma^{(0)}}$ exists such that $Q_c(\gamma^{(0)})$ and $Q_c^\prime(\gamma^{(0)})$ are well defined.
This also shows that in every iterative step, $\gamma^{(k+1)}$ exists as long as $\gamma^{(k)}$ exists.

Recall the definition of $Q_c(\gamma)$ and $Q(\beta, \gamma)$ in \eqref{definition-Q} and \eqref{definition-Qc}.
Note that $\kappa_n=\max_{i,j} \|z_{ij}\|_\infty$. By Condition \ref{condition-Q} and inequality \eqref{ineq-mu-beta}, we have
\begin{eqnarray*}
\|Q_c(\gamma^*)\|_\infty & \le & \|\sum_{i<j} z_{ij}( a_{ij} - \E a_{ij})\|_\infty + \|\sum_{i< j} z_{ij}[\mu_{ij}(\beta^*, \gamma^*)-  \mu_{ij}(\widehat{\beta}_{\gamma^*}, \gamma^*) ]\|_\infty \\
& = & O( h_{n2}n^{3/2}(\log n)^{1/2} ) + O( n^2 \kappa_n b_{n1} \times \frac{b_{n1}^2(h_{n1}+b_{n1}\kappa_n)}{b_{n0}^3} \sqrt{ \frac{\log n}{n} }  ) \\
& = & O( [h_{n2} +\kappa_n b_{n1}^3b_{n0}^{-3} (h_{n1}+b_{n1}\kappa_n) ]n^{3/2}(\log n)^{1/2} ). 
\end{eqnarray*}
By Proposition \ref{proposition-Q-Lip}, $\lambda=n^2 \kappa_n^4 b_{n1}^{11} b_{n2} b_{n0}^{-9} $.
Let $\eta_n= h_{n2} +\kappa_n b_{n1}^3 (h_{n1}+b_{n1}\kappa_n)b_{n0}^{-3}$.
By Condition \ref{condition-H},
$\aleph=\| [Q_c^\prime(\gamma^*)]^{-1} \|_\infty = O (h_{n3}n^{-2})$.
Thus,
\[
\delta = \| [Q_c^\prime(\gamma^*)]^{-1} Q_c(\gamma^*) \|_\infty = O( \eta_n h_{n3} n^{-1/2}(\log n)^{1/2} ).
\]
 As a result, if equation \eqref{eq:consistency} holds, then
\[
\rho=2\aleph \lambda \delta = O(\eta_n h_{n3}^2  \kappa_n^4 b_{n1}^{11} b_{n2} b_{n0}^{-9} n^{-1/2}(\log n)^{1/2})=o(1).
\]
By Theorem \ref{pro:Newton:Kantovorich}, with probability approaching one, the limiting point of the sequence $\{\gamma^{(k)}\}_{k=1}^\infty$ exists denoted by $\widehat{\gamma}$ and satisfies
\[
\| \widehat{\gamma} - \gamma^* \|_\infty \le \delta.
\]
At the same time, by Lemma \ref{lemma-a}, $\widehat{\beta}_{\widehat{\gamma}}$ exists and $(\widehat{\gamma}, \widehat{\beta}_{\widehat{\gamma}})$ is the moment estimator.
It completes the proof.

\subsection{Proof of Theorem~\ref{theorem-central-b}}

Write $\widehat{\beta}^*=\widehat{\beta}(\gamma^*)$.
Recall that $T_{ij}$ is a vector of length $n$ with $i$th and $j$th elements ones and other elements zeros and
\[
s_{\beta_{ij}}(\beta, \gamma)= (a_{ij} - \mu_{ij}(\beta, \gamma))T_{ij}, ~~s_{\gamma_{ij}}(\beta, \gamma)= z_{ij}(a_{ij} - \mu_{ij}(\beta, \gamma)),
\]
\[
V(\beta, \gamma)= \frac{ \partial F(\beta, \gamma)}{\partial \beta},~~V^*=V(\beta^*, \gamma^*).
\]
To show Theorem~\ref{theorem-central-b}, we need one lemma below, whose proof is in the supplementary material.

\begin{lemma}
\label{lemma-thereom-gamma-a}
Under conditions \ref{condition-diff-a} -- \ref{condition-H}, if \eqref{equation:lemma-a} and the following hold:
\begin{equation}\label{condition-lemma2}
\frac{b_{n2} b_{n1}^6\log n(h_{n1}+b_{n1})^2}{n^{1/2}b_{n0}^9} )=o(1),
\end{equation}
then
\begin{equation}\label{result-lemma2}
\sqrt{n}(\widehat{\beta}^* - \beta^*) = - (\frac{1}{n} V^*)^{-1}  \sum_{j<i} s_{\beta_{ij}}(\beta^*, \gamma^*) + o_p(1).
\end{equation}
\end{lemma}

Now we give the proof of Theorem \ref{theorem-central-b}.

\begin{proof}[Proof of Theorem \ref{theorem-central-b}]
Recall that $F(\widehat{\beta}(\gamma), \gamma)=0$, where $F(\cdot, \cdot)$ is defined at equation \eqref{eqn:def:F}, and $\widehat{\beta}=\widehat{\beta}(\widehat{\gamma})$.
A mean value expansion gives
\[
 \sum_{j< i} s_{\gamma_{ij}}( \widehat{\beta} ,  \widehat{\gamma} )
-  \sum_{j< i} s_{\gamma_{ij}}(\widehat{\beta}(\gamma^*), \gamma^*  )
=   \sum_{j< i} \frac{\partial }{ \partial \gamma^\top } s_{\gamma_{ij}} (
\widehat{\beta}(\bar{\gamma}), \bar{\gamma}) (\widehat{\gamma}-\gamma^*),
\]
where $\bar{\gamma}=t\gamma^*+(1-t)\widehat{\gamma}$ for some $t\in (0, 1)$.
By noting that
\[
Q(\widehat{\beta}, \widehat{\gamma}) = \sum_{j< i} s_{\gamma_{ij}}(\widehat{\beta}, \widehat{\gamma})=0,
\]
we have
\[
\sqrt{N}(\widehat{\gamma} - \gamma^*) = -
\Big[ \frac{1}{N}  \sum_{j< i} \frac{\partial }{ \partial \gamma^\top } s_{\gamma_{ij}} (\widehat{\beta}(\bar{\gamma}), \bar{\gamma}) \Big]^{-1}
\times \Big[\frac{1}{\sqrt{N}}  \sum_{j< i} s_{\gamma_{ij}}(\widehat{\beta}(\gamma^*), \gamma^*) \Big].
\]
Note that the dimension of $\gamma$ is fixed. By Theorem \ref{Theorem:con} and \eqref{equation-H-appro}, we have
\[
-\frac{1}{N}  \sum_{j< i} \frac{\partial}{ \partial \gamma^\top } s_{\gamma_{ij}}
(\widehat{\beta}(\bar{\gamma}), \bar{\gamma}  )
\stackrel{p}{\to } \bar{H}:= \frac{1}{N}H(\beta^*, \gamma^*).
\]
Write $\widehat{\beta}^*$ as $\widehat{\beta}(\gamma^*)$ for convenience. Therefore,
\begin{equation}\label{eq:theorem4:aa}
\sqrt{N} (\widehat{\gamma} - \gamma^*) = \bar{H}^{-1} \Big[ \frac{1}{\sqrt{N}}  \sum_{j< i}
s_{\gamma_{ij}} ( \widehat{\beta}^*, \gamma^*   )\Big] + o_p(1).
\end{equation}
By applying a third order Taylor expansion to the summation in brackets in \eqref{eq:theorem4:aa}, it yields
\begin{equation}\label{eq:gamma:asym:key}
\frac{1}{\sqrt{N}}  \sum_{j< i} s_{\gamma_{ij}} (\widehat{\beta}^*, \gamma^*) = S_1 + S_2 + S_3,
\end{equation}
where
\begin{equation*}
\begin{array}{l}
S_1  =  \frac{1}{\sqrt{N}}  \sum_{j< i} s_{\gamma_{ij}} (\beta^*, \gamma^* )
+ \frac{1}{\sqrt{N}}  \sum_{j< i}
\Big[\frac{\partial}{\partial \beta^\top } s_{\gamma_{ij}} (\beta^*, \gamma^*)\Big]( \widehat{\beta}^* - \beta^* ), \\
S_2  =   \frac{1}{2\sqrt{N}} \sum_{k=1}^{n} \Big[( \widehat{\beta}_k^* - \beta_k^* ) \sum_{j< i}
\frac{\partial^2 }{ \partial \beta_k \partial \beta^\top } s_{\gamma_{ij}} (\beta^*, \gamma^* )
\times ( \widehat{\beta}^* - \beta^* ) \Big],  \\
S_3  =  \frac{1}{6\sqrt{N}} \sum_{k=1}^{n} \sum_{l=1}^{n} \{ (\widehat{\beta}_k^* - \beta_k^*)(\widehat{\beta}_l^* - \beta_l^*)
\Big[  \sum_{j< i} \frac{ \partial^3 s_{\gamma_{ij}} (\bar{\beta}^*, \gamma^*)}{ \partial \beta_k \partial \beta_l \partial \beta^\top } \Big]
(\widehat{\beta}^*  - \beta^* )\},
\end{array}
\end{equation*}
and $\bar{\beta}^*=t\beta^*+(1-t)\widehat{\beta}^*$ for some $t\in(0,1)$.
Similar to the proof of Theorem 4 in \cite{Graham2017}, we will show that
(1) $S_2$ is the bias term having a non-zero probability limit; (2) $S_3$ is an asymptotically negligible
remainder term.

We first evaluate the term $S_3$.
We calculate $g^{ij}_{klh}=\frac{ \partial^3 s_{\gamma_{ij}} (\beta, \gamma ) }{ \partial \beta_k \partial \beta_l \partial \beta_h }$ according to the indices $k,l,h$ as follows.
We first observe that $g^{ij}_{klh}=0$ when $k,l,h\notin \{i,j\}$ since $s_{\gamma_{ij}} (\beta, \gamma )$ only has the arguments $\beta_i$ and $\beta_j$ in regardless of other $\beta_k$'s ($k\neq i, j$).
So there are only two cases below in which  $g^{ij}_{klh}\neq 0$. \\
(1) Only two values among three indices $k, l, h$ are equal.
If $k=l=i; h=j$,
$g^{ij}_{klh}=-z_{ij}\frac{\partial^3 \mu_{ij}}{\partial \pi_{ij}^3}$; for other cases, the results are similar.\\
(2) Three values are equal.
$g^{ij}_{klh}=-z_{ij}\frac{\partial^3 \mu_{ij}}{\partial \pi_{ij}^3}$ if $k=l=h=i$ or $k=l=h=j$. \\
Therefore, we have
\begin{eqnarray*}
S_3&=&\frac{1}{6\sqrt{N}}  \sum_{j < i} \sum_{k, l, h}
\frac{ \partial^3 s_{\gamma_{ij}} (\bar{\beta}^*, \gamma^*)}{ \partial \beta_k \partial \beta_l \partial \beta_h }(\widehat{\beta}_k^* - \beta_k^*)(\widehat{\beta}_l^* - \beta_l^*)
(\widehat{\beta}_h^* - \beta_h^*) \\
& = & \frac{1}{6\sqrt{N}}  \sum_{j < i} \left \{
 3\frac{ \partial^3 s_{\gamma_{ij}} (\bar{\beta}^*, \gamma^*)}{ \partial \beta_i^2 \partial \beta_j}(\widehat{\beta}_i^* - \beta_i^*)^2(\widehat{\beta}_j^* - \beta_j^*)
 + 3 \frac{ \partial^3 s_{\gamma_{ij}} (\bar{\beta}^*, \gamma^*)}{ \partial \beta_j^2 \partial \beta_i}(\widehat{\beta}_j^* - \beta_j^*)^2(\widehat{\beta}_i^* - \beta_i^*)
 \right.
 \\
 &&
 \left.
 +\frac{ \partial^3 s_{\gamma_{ij}} (\bar{\beta}^*, \gamma^*)}{ \partial \beta_i^3 }(\widehat{\beta}_i^* - \beta_i^*)^3
 +\frac{ \partial^3 s_{\gamma_{ij}} (\bar{\beta}^*, \gamma^*)}{ \partial \beta_j^3 }(\widehat{\beta}_j^* - \beta_j^*)^3
 \right \}.
\end{eqnarray*}
So
\begin{eqnarray*}
\|S_3\|_\infty & \le &\frac{4}{3\sqrt{N}} \times \max_{i,j} \left\{|\frac{\partial^3 \mu_{ij}(\bar{\beta}^*, \gamma^*))}{\partial \pi_{ij}^3}| \| z_{ij} \|_\infty \right\}
\times \frac{n(n-1)}{2} \| \widehat{\beta}^* - \beta\|_\infty^3.
\end{eqnarray*}
By Lemma \ref{lemma-a}  and inequality \eqref{ineq-mu-beta}, we have
\[
\|S_3\|_\infty =  O_p( \frac{b_{n3}\kappa_n b_{n1}^6(h_{n1} + b_{n1}\kappa_n )^3(\log n)^{3/2}}{n^{1/2}b_{n0}^9} ).
\]

Similar to the calculation in the derivation of the asymptotic bias in Theorem 4 in \cite{Graham2017}, we have
$S_2=B_*+o_p(1)$, where $B_*$ is defined at \eqref{defintion-Bias}.

By Lemma \ref{lemma-thereom-gamma-a}, similar to the calculation in the derivation of the asymptotic expression of $S_1$ in \cite{Graham2017}, we have
\[
S_1 = \frac{1}{\sqrt{N}} \sum_{j< i} \tilde{s}_{\gamma_{ij}} (\beta^*, \gamma^*) + o_p(1),
\]
Therefore, it shows that equation \eqref{eq:gamma:asym:key} is equal to
\begin{equation}\label{eq:proof:4-a}
\frac{1}{\sqrt{N}} \sum_{j< i} s_{\gamma_{ij}}( \widehat{\beta}^*, \gamma^* )
= \frac{1}{\sqrt{N}} \sum_{j< i} \tilde{s}_{\gamma_{ij}}( \beta^*, \gamma^* ) + B_* + o_p(1),
\end{equation}
with $\frac{1}{\sqrt{N}} \sum_{i=1}^n \sum_{j\neq i} s_{\gamma_{ij}}^*( \beta^*, \gamma^* )$ equivalent to the first two terms in \eqref{eq:gamma:asym:key}
and $B_*$ the probability limit of the third term in \eqref{eq:gamma:asym:key}.

Substituting \eqref{eq:proof:4-a} into \eqref{eq:theorem4:aa} then gives
\[
\sqrt{N}(\widehat{\gamma}- \gamma^*) = \bar{H}^{-1} B_* + \bar{H}^{-1} \times \frac{1}{\sqrt{N}}  \sum_{j< i}
s_{\gamma_{ij}}^* (\beta^*, \gamma^*) + o_p(1).
\]
It completes the proof.
\end{proof}

\subsection{Proofs for Theorem \ref{Theorem-central-a}}


To simplify notations, write  $\mu_{ij}^\prime = \mu^\prime(\beta_i^* + \beta_j^* + z_{ij}^\top \gamma^*)$ and
\[
V= \frac{ \partial F(\beta^*, \gamma^*)}{\partial \beta^\top}, ~~ V_{\gamma\beta} = \frac{ \partial F(\beta^*, \gamma^*)}{\partial \gamma^\top}.
\]
Recall that $\pi_{ij}=\beta_i+\beta_j+z_{ij}^\top \gamma$.
By a second Taylor's expansion, we have
\begin{equation}
\label{equ-Taylor-exp}
\mu( \widehat{\beta}_i+\widehat{\beta}_j + z_{ij}^\top \widehat{\gamma}) - \mu(\beta_i^*+\beta_j^* + \gamma^*)
= \mu_{ij}^\prime (\widehat{\beta}_i-\beta_i)+\mu_{ij}^\prime (\widehat{\beta}_j-\beta_j) + \mu_{ij}^\prime z_{ij}^\top ( \widehat{\gamma} - \gamma)
+ g_{ij},
\end{equation}
where
\[
g_{ij}= \frac{1}{2} (\widehat{\beta}_i-\beta_i^*, \widehat{\beta}_j-\beta_j^*,  ( \widehat{\gamma} - \gamma^* )^\top  )
\begin{pmatrix}
\mu^{\prime\prime}( \tilde{\pi}_{ij} ) & \mu^{\prime\prime}( \tilde{\pi}_{ij} )  & \mu^{\prime\prime}( \tilde{\pi}_{ij} ) z_{ij}^\top \\
\mu^{\prime\prime}( \tilde{\pi}_{ij} ) & \mu^{\prime\prime}( \tilde{\pi}_{ij} )  & \mu^{\prime\prime}( \tilde{\pi}_{ij} ) z_{ij}^\top \\
\mu^{\prime\prime}( \tilde{\pi}_{ij} ) z_{ij}^\top & \mu^{\prime\prime}( \tilde{\pi}_{ij} ) z_{ij}^\top & \mu^{\prime\prime}( \tilde{\pi}_{ij} ) z_{ij}z_{ij}^\top
\end{pmatrix}
\begin{pmatrix}
\widehat{\beta}_i-\beta_i^* \\
\widehat{\beta}_j-\beta_j^* \\
\widehat{\gamma} - \gamma^*
\end{pmatrix},
\]
and $\tilde{\pi}_{ij}$ lies between $\pi_{ij}^*$ and $\widehat{\pi}_{ij}$.
By calculations, $g_{ij}$ can be simplified as
\begin{eqnarray*}
g_{ij} & = &  \mu^{\prime\prime}( \tilde{\pi}_{ij} ) [(\widehat{\beta}_i-\beta_i)^2 +  (\widehat{\beta}_j-\beta_j)^2 + 2(\widehat{\beta}_i-\beta_i)(\widehat{\beta}_j-\beta_j)]
\\
&& + 2\mu^{\prime\prime}( \tilde{\pi}_{ij} ) z_{ij}^\top ( \widehat{\gamma} - \gamma) (\widehat{\beta}_i-\beta_i+\widehat{\beta}_j-\beta_j) +
( \widehat{\gamma} - \gamma)^\top  \mu^{\prime\prime}( \tilde{\pi}_{ij} ) z_{ij}z_{ij}^\top( \widehat{\gamma} - \gamma)
\end{eqnarray*}
Note that $\kappa_n := \max_{i,j} \| z_{ij} \|$ and $|\mu^{\prime\prime}(\pi_{ij})|\le b_{n2}$ when $\beta \in B(\beta^*, \epsilon_{n1})$ and $\gamma \in B(\gamma^*, \epsilon_{n2})$.
So we have
\begin{equation}
\label{inequality-gij}
\begin{array}{rcl}
|g_{ij}| & \le & 4b_{n2} \| \widehat{\beta} - \beta^*\|_\infty^2 + 2b_{n2}\| \widehat{\beta} - \beta^*\|_\infty \| \widehat{\gamma}-\gamma^* \|_1 \kappa_n + b_{n2} \| \| \widehat{\gamma}-\gamma^* \|_1^2 \kappa_n^2 \\
& \le & 2b_{n2}[4 \| \widehat{\beta} - \beta^*\|_\infty^2+  \| \widehat{\gamma}-\gamma^* \|_1^2 \kappa_n^2].
\end{array}
\end{equation}
Let  $g_i=\sum_{j\neq i}g_{ij}$, $g=(g_1, \ldots, g_n)^\top$.
By equation \eqref{equ-Taylor-exp}, we have
\[
d - \E d = V(\widehat{\beta} - \beta^*) + V_{\gamma\beta} (\widehat{\gamma}-\gamma^*) + g.
\]
Equivalently,
\begin{equation}
\label{expression-beta}
\widehat{\beta} - \beta^* = V^{-1}(d - \E d) + V^{-1}V_{\gamma\beta} (\widehat{\gamma}-\gamma^*) + V^{-1} g.
\end{equation}

Now we evaluate the last remainder term in the right hand of the above equation.
Let $W=V^{-1} - S$.
Note that $(S g)_i =  g_i/v_{ii}$ and $(n-1)b_{n0} \le v_{ii} \le (n-1)b_{n1}$.
By inequality \eqref{inequality-gij} and Theorem \ref{Theorem:con}, we have
\begin{eqnarray*}
\max_i |(S g)_i| & \le & \max_i \frac{|g_i|}{\hat{v}_{ii}}  \le  \frac{2b_{n2}}{b_{n0}}\left(4 \| \widehat{\beta} - \beta^*\|_\infty^2+  \| \widehat{\gamma}-\gamma^* \|_1^2 \kappa_n^2\right) \\
& = & O_p\left( \frac{b_{n2}}{b_{n0}} (\varphi_{n1}^2+\varphi_{n2}^2\kappa_n^2) \frac{\log n}{n} \right),
\end{eqnarray*}
where
\[
\varphi_{n1}=b_{n1}^2b_{n0}^{-3}  (h_{n1}+b_{n1}\kappa_n), ~~~ \varphi_{n2}= h_{n3}[h_{n2} + \kappa_n b_{n1}^3 (h_{n1}+ b_{n1}\kappa_n)b_{n0}^{-3}].
\]
By Proposition \ref{pro:inverse:appro}, we have
\begin{equation*}
 \|W g \|_\infty \le  n\|R \|_\infty \|g\|_\infty
= O_p\left( \frac{b_{n1}^2(\varphi_{n1}^2+\varphi_{n2}^2\kappa_n^2)b_{n2}}{b_{n0}^3} \frac{\log n}{n} \right).
\end{equation*}
So if
\[
\frac{b_{n1}^2(\varphi_{n1}^2+\varphi_{n2}^2\kappa_n^2)b_{n2}}{b_{n0}^3} \frac{\log n}{n^{1/2}}=o(1),
\]
then
\begin{equation}
\label{equ-thoerem3-2}
\|V^{-1} g\|_\infty = o_p(n^{-1/2}).
\end{equation}

Note that
\[
V_{\gamma\beta} = \left(\begin{array}{c}
\sum_{j=1,j\neq 1}^n \mu^\prime_{1j} z_{1j}^\top, \\
\vdots \\
\sum_{j=1, j\neq n}^n \mu^\prime_{nj} z_{nj}^\top,
\end{array}
\right).
\]
So we have
\[
\| V_{\gamma\beta}(\widehat{\gamma}-\gamma^*) \|_\infty \le (n-1) \kappa_n\|\widehat{\gamma}-\gamma^*\|_1.
\]
By Proposition \ref{pro:inverse:appro}, we have
\begin{equation}
\label{equ-theorem3-3}
\begin{array}{rcl}
\|V^{-1}V_{\gamma\beta} (\widehat{\gamma}-\gamma^*)\|_\infty   & \le & \max_i \frac{1}{v_{ii}}  \|V_{\gamma\beta} (\widehat{\gamma}-\gamma^*)\|_\infty
+ n \|W\|_{\max} \|V_{\gamma\beta} (\widehat{\gamma}-\gamma^*)\|_\infty \\
& \le & O_p\left( \frac{b_{n1}^2\kappa_n\|(\widehat{\gamma}-\gamma^*)\|_1}{b_{n0}^3} \right).
\end{array}
\end{equation}
Note that $R=\mathrm{Cov}(d)$ and $\mathrm{Cov} [W(d - \E d)] = W R W^\top$. If
\[
\max_i |(WBW^\top)_{ii}| \le \|W\|_{\max}^2 \sum_{i,j}|R_{ij}| = \frac{b_{n1}^4}{n^4b_{n0}^4} \sum_{i,j}|R_{ij}|= o( \frac{1}{n} ),
\]
then
\begin{equation}
\label{equ-theorem3-dd}
\P( [W(d - \E d)]_i > o(n^{-1/2}) ) \le O(n) |\mathrm{Var}\{[W(d - \E d)]_i\}| =o(1).
\end{equation}
Consequently, by combining \eqref{expression-beta}, \eqref{equ-thoerem3-2}, \eqref{equ-theorem3-3} and \eqref{equ-theorem3-dd},
we have
\[
\widehat{\beta}_i - \beta^*_i = [S(d - \E d)]_i + [W(d - \E d)]_i + o_p(n^{-1/2})=[S(d - \E d)]_i+ o_p(n^{-1/2}).
\]
It completes the proof.
